\documentclass{aa}
\usepackage{graphics}
\newcommand{\gsim}{\,\raisebox{0.2em}{$>$}\!\!\!\!\!
\raisebox{-0.25em}{$\sim$}\,}
\newcommand{\lsim}{\,\raisebox{0.2em}{$<$}\!\!\!\!\!
\raisebox{-0.25em}{$\sim$}\,}
\newcommand{\gr}{$\gamma$-ray \,}

\begin{document}

\thesaurus{08(02.19.1, 08.19.4, 09.03.2, 09.19.2, 13.07.3)}
%


\title{ Kinetic theory of cosmic ray and gamma-ray production\\
 in supernova remnants expanding into wind bubbles}

\author{Evgeny G.\, Berezhko\inst{1} 
 \and Heinrich J.\, V\"olk\inst{2}}

\offprints{H.J. V\"olk} 

\institute{Institute of Cosmophysical Research
 and Aeronomy, Lenin Ave. 31, 677891
Yakutsk, Russia\\ email:
 berezhko@sci.yakutia.ru 
 \and Max-Planck-Institut f\"ur Kernphysik,
 Postfach 103980, D-69029 Heidelberg, Germany\\ email:
 Heinrich.Voelk@mpi-hd.mpg.de}

\date{Received <03 December 1999> / Accepted <date>}

\titlerunning{CRs and gamma-rays in wind SNRs}

\authorrunning{E.G. Berezhko \& H.J. V\"olk}

\maketitle

\begin{abstract} 
A kinetic model of particle acceleration in supernova remnants (SNRs) is
extended to study the cosmic ray (CR) and associated high-energy
$\gamma$-ray production during SN shock propagation through the
inhomogeneous circumstellar medium of a progenitor star that emits a wind.
The wind forms a low-density bubble surrounded by a swept-up shell of
interstellar matter. $\gamma$-rays are produced as a result of decay of
pions which in turn are the result of collisions of CRs with nuclei of the
thermal plasma. The time evolution of the SNRs is followed numerically,
taking into account the nonlinear backreaction of the accelerated CRs. The
model for SNRs includes injection of suprathermal particles at the shock
front and heating of the thermal plasma due to dissipation of Alfv\'en
waves in the precursor region. Examples typical for SN type Ib and SN type
II explosions are considered. Apart from the confirmation of the known
fact that acceleration is extremely rapid and that the upper momentum
cutoff is reached almost immediately after the explosion due to the high
wind magnetic field, it is also shown that the CRs are accelerated with a
high efficiency. Depending on the circumstellar parameters, 20\% to 40\%
of the SN explosion energy is absorbed by CRs during the SNR evolution for
a moderate injection rate, when a fraction $\eta = 10^{-3}$ of the gas
particles, swept up by the supernova shock, is accelerated. The CR
momentum spectrum, ultimately produced in the SNRs, has a power law form
$N\propto p^{-\gamma}$ with an index $\gamma=2.0$ to 2.1 in a wide energy
range up to a maximum energy, which is about $10^{14}$~eV, if the CR
diffusion coefficient is as small as the Bohm limiting value. It is to be
expected
that the resulting CR chemical composition at high energies reflects more
the stellar wind composition, whereas at lower energies it corresponds
more to that of the average interstellar medium. The expected
$\pi^0$-decay $\gamma$-ray flux is however considerably lower than in the
case of a uniform interstellar medium; a relatively high $\gamma$-ray
luminosity in the band $\epsilon_{\gamma}\gsim 1$~TeV, detectable at
distances of several kpc, is only expected in the case of a relatively
dense ISM with a number density above 10~cm$^{-3}$.  Extremely high
$\gamma$-ray emission may be produced when the SN shock propagates through
the slow dense wind of a red supergiant, the progenitor of a SN type II.
In this case SNRs might be visible in $\gamma$-rays for several hundred
years out to distances of tens of kpc. For the case of a SN type Ib the
expected $\pi^0$-decay {\gr} TeV-energy flux during the whole SNR
evolution remains lower than $10^{-11}$~cm$^{-2}$s$^{-1}$ if the
interstellar number density is less than $0.3$~cm$^{-3}$.
\end{abstract}

\section{Introduction} 

Considerable efforts have been made during the last years to empirically
confirm the theoretical expectation that the main part of the Galactic
cosmic rays (CRs) originates in supernova remnants (SNRs). Theoretically
progress in the solution of this problem has been due to the development
of the theory of diffusive shock acceleration (see, for example, reviews
by Drury 1983; Blandford \& Eichler 1987; Berezhko \& Krymsky 1988).
Although still incomplete, the theory is able to explain the main
characteristics of the observed CR spectrum under several reasonable
assumptions, at least up to an energy of $10^{14}\div 10^{15}$~eV. Direct
information about the dominant nucleonic CR component in SNRs can only be
vobtained from $\gamma$-ray observations. If this nuclear component is
strongly enhanced inside SNRs then through inelastic nuclear collisions,
leading to pion production and subsequent decay, $\gamma$-rays will be
produced.

CR acceleration in SNRs expanding in a uniform interstellar medium (ISM)
(Drury et al. 1989; Markiewicz et al. 1990; Dorfi 1990), and the
properties of the associated $\gamma$-ray emission (Dorfi 1991; Drury et
al. 1994) were investigated in a number of studies (we mention here only
those papers which include the effects of shock geometry and
time-dependent nonlinear CR backreaction; for a review of others which
deal with the test particle approximation, see for example Drury 1983;
Blandford \& Eichler 1987; Berezhko \& Krymsky 1988; V\"olk 1997). All of
these studies are based on a two-fluid hydrodynamical approach and
directly employ the assumption that the expanding SN shock is locally
plane; as dynamic variables for the CRs the pressure and the energy
density are determined. Their characteristics are sometimes essentially
different from the results obtained in a kinetic approach (Kang \& Jones
1991; Berezhko et al. 1994, 1995, 1996) which consistently takes the role
of shock geometry and nonlinear CR backreaction into account. First of
all, in kinetic theory the form of the spectrum of accelerated CRs and
their maximum energy are calculated selfconsistently. In particular, the
maximum particle energy $\epsilon_{max}$, achieved at any given
evolutionary stage, is determined by geometrical factors (Berezhko 1996),
in contrast to the hydrodynamic models which in fact postulate that the
value of $\epsilon_{max}(t)$ is determined by the time interval $t$ that
has passed since the explosion (Drury et al. 1989; Markiewicz et al. 1990;
Dorfi 1990; Jones \& Kang 1992). Although the difference between the
values of $\epsilon_{max}$ in the two cases is not very large, it
critically influences the structure and evolution of the shock. For
example, the shock never becomes completely modified (smoothed) by the CR
backreaction (Kang \& Jones 1991; Berezhko et al. 1994, 1995, 1996).
Together with the smooth precursor, the shock transition always contains a
relatively strong subshock which heats the swept-up gas and leads to the
injection of suprathermal gas particles into the acceleration process. In
this sense diffusive shock acceleration is somewhat less efficient than
predicted by hydrodynamic models. Acceleration always requires some
freshly injected particles which are generated during gas heating. This
prediction is in agreement with the observations that show significant gas
heating in young SNRs.

On the other hand, the shock modification by the CR backreaction is even
greater than predicted by hydrodynamic models. The total shock compression
ratio $\sigma \propto M^{3/4}$ (Berezhko et al. 1996) is a monotonically
rising function of the shock Mach number $M$ and can significantly exceed
the classical value 4, or even the value $\sigma =7$ that corresponds to a
postshock medium dominated by relativistic CRs. This result is a direct
consequence of the fact that CR acceleration at the front of a spherically
expanding shock is accompanied by a temporally increasing dilution of
shock energy in the form of CRs in the precursor region (Drury et al.
1995; Berezhko 1996). Qualitatively this leads to the same effect as if
energetic particles or thermal gas energy were radiated away from the
shock (Berezhko 1996; Berezhko \& Ellison 1999). We note that this energy
dilution effect is much stronger than the effect of a decreasing specific
heat ratio due to the conversion of shock energy into relativistic CRs.

The additional gas heating in the precursor region due to Alfv\'en wave
dissipation significantly restricts the shock compression ratio
$\sigma(t)$ even though its value still considerably exceeds 4 for a
strong shock (Berezhko et al. 1996; Berezhko \& V\"olk 1997). The
precursor gas heating has a much less pronounced effect in the
hydrodynamic description (e.g. Dorfi 1990).

As far as the expected $\gamma$-ray emission, produced in SNRs by the
nuclear CR component, is concerned, there are less significant differences
between the kinetic (Berezhko \& V\"olk 1997) and the hydrodynamic (Dorfi
1991; Drury et al. 1994) predictions, even though these differences are
not unimportant. The reason is that the peak value of the $\gamma$-ray
flux is mainly determined by the fraction of overall hydrodynamic
explosion energy that is transferred to CRs, i.e. by the overall
efficiency of CR acceleration which is not strongly dependent on the model
used.

The differences in the time variation of the predicted $\gamma$-ray fluxes
are more essential. Kinetic theory (Berezhko \& V\"olk 1997) revealed much
more effective CR and $\gamma$-ray production during the free expansion
phase. It also shows a more rapid decrease of the $\gamma$-ray flux during
the subsequent Sedov phase after reaching its peak value, due to the
different spatial distributions of thermal gas and CRs inside SNRs. This
energy-dependent lack of overlap is not taken into account in
hydrodynamic models.

The model developed by Ellison and co-workers to describe diffusive shock
acceleration of CRs and to predict the expected $\gamma$-ray emission
from SNRs, uses a Monte Carlo (MC) simulation of pitch angle scattering.
It is applied in the same way to all particles, both CR and thermal
particles, and is correspondingly solved numerically from the outset
(e.g. Ellison et al. 1995; Baring et al. 1999). The energetic particle
population (i.e. accelerated CRs) naturally arises in this approach as a
high energy tail of the distribution of gas particles which undergo
heating in the strongest gradient of the selfconsistently modified shock
transition, the subshock in our notation. In this way the model
incorporates not only selfconsistent CR acceleration but also gas heating
and particle injection into the acceleration process. Since it is a
plane-wave steady state model it includes as a free parameter the value
of the CR cutoff energy $\epsilon_{max}$, which can not be calculated in
this kind of model. On the other hand, at sufficiently high particle
energies the description corresponds to a diffusive approach based on the
diffusive transport equation. Therefore our time-dependent kinetic
description in spherical symmetry, with an injection rate that
corresponds to the MC-model, and the MC-model, in turn incorporating the
same value of CR cutoff energy, give identical CR spectra at any stage of
the free expansion phase for the same set of ISM and shock parameters
(Berezhko \& Ellison 2000). This means that in its free expansion phase
the SNR evolution can be represented as a sequence of quasi-stationary
states; each of them can be reproduced in the framework of the plane-wave
steady state description with the corresponding value of the CR cutoff
energy $\epsilon_{max}(t)$. The situation becomes more complicated in the
subsequent Sedov phase due to the existence of a so-called escaping
particle population near the shock front (Berezhko \& Ellison 1999b),
which is a purely nonstationary phenomenon (Berezhko 1986a; Berezhko \&
Krymsky 1988) and therefore can not be reproduced in the framework of a
steady state approach. As far as the overall CR spectrum and the
associated $\gamma$-ray flux at any given evolutionary phase are
concerned, the plane wave description can only give an approximate
estimate because it does not include such important physical factors as
CR adiabatic cooling and the incomplete overlap of the spatial
distributions of gas and CRs inside the expanding SNR.

The evolution of SNRs in a uniform ISM is typical only for SNe type Ia.
SNe of type Ib and II, which are more numerous in our Galaxy, explode
into an inhomogeneous circumstellar me\-di\-um, formed by the intensive
wind of their massive progenitor stars (Losinskaya 1991). This is at
least true for progenitor masses $M \gsim 20 M_{\odot}$ whose main
sequence winds are very energetic. Their time-integrated hydrodynamic
energy output is so large that they create wind cavities which are about
as large as the size of a remnant of the subsequent Supernova explosion,
if it would occur in a uniform ISM of the same density. The more frequent
lower
mass core collapse progenitors ($8 \lsim M \l 20 M_{\odot}$) have at
least a massive Red Supergiant wind region before the transition to the
external ISM. Therefore the description of SNR evolution and CR
acceleration becomes considerably more difficult for those cases. Due to
the complicated structure of the medi\-um one can expect that the
difference between kinetic and hydrodynamic model predictions will be
even more pronounced than in the case of a uniform ISM.

Up to now, except for the paper by Berezhko \& V\"olk (1995) where
preliminary results of the present work were presented, the consideration
of CR acceleration in the case of a SN type Ib and SN type II was
considered only for SN shock propagation through the region of the
supersonic stellar wind (Berezinskii \& Ptuskin 1988; V\"olk \& Biermann
1988; Jones \& Kang 1992; Kirk et al. 1995). Since this region is
characterized by an initially very strong if spatially decreasing magnetic
field of stellar origin and a very high shock speed, the maximum energy of
CRs accelerated in the wind region is achieved extremely rapidly during a
period of time comparable with the SNR age at each phase, and then it
remains nearly constant due to a balance of acceleration gains and
adiabatic losses (V\"olk \& Biermann 1988). As it is shown below, the
expected CR energy spectrum is close to being selfsimilar where the
amplitude monotonically decreases with time but the shape remains
invariant (see Appendix). Thus the situation is very different from that
for a uniform circumstellar medium with a standard magnitude ISM field
strength, where the maximum CR energy is only achieved at the end of the
sweep-up phase, after hundreds or even thousands of years. 

The direct dependence on the stellar surface field also raised hopes that
the CR acceleration in the wind would constitute a major source of the
observed Galactic CRs with energies $\epsilon \gsim 10^{14}$~eV. The most
optimistic assumptions predict CR generation in these regions up to an
energy of $3\times 10^{18}$~eV (Biermann 1993). In this context we would
like to make the following remarks:

First of all, the supersonic wind region contains a relatively small
amount of mass $M\lsim 1 M_{\odot}$ which is typically smaller than the
ejected mass $M_{ej}\sim 10M_{\odot}$. Therefore the fraction of the SN
explosion energy which can be transfered into CRs is typically lower than
required for the observed CR spectrum, especially in the case of a SN type
Ib (see below).

Second, to achieve a particle energy significantly greater than
$10^{14}$~eV one needs to assume an unusually high magnetic field in the
wind. This tends to violate the condition $V_{w} \geq c_{a}$, necessary
for the existence of a supersonic flow in the first place, as pointed out 
by Axford (1994), where $V_w$ is the wind speed and $c_a$ is the Alfv\'en
velocity.

Third, the wind magnetic field is largely azimuthal, except in the very
polar region near the stellar rotation axis, and the outer SNR shock is
almost perpendicular. Therefore not only the suprathermal ion injection is
much less efficient but also the CR acceleration itself appears not so
efficient as in the case of a quasi-parallel shock. Although one can give
some physical arguments which assert that particle injection with their
subsequent acceleration takes place also in this case, one needs a more
rigorous treatment of this complex question. An example for the
acceleration at a shock propagating through a stationary stochastic large
yyscale magnetic field that is on average perpendicular to the shock
normal
is given by Kirk et al. (1996) based on the process of anomalous diffusion
(see Chuvilgin \& Ptuskin 1993). This test particle calculation indicates,
as intuitively expected, that the spectrum of accelerated particles is
steeper than for acceleration at a quasi-parallel shock. As a consequence,
the nonlinear efficiency of acceleration would also have to be assumed to
be significantly reduced.

Fourth, the unusually high maximum CR energy predicted by Biermann (1993)  
is a direct consequence of the assumption of a very small and energy
independent CR
diffusion coefficient, presumably produced by the chaotic turbulent
motions of the medium near the shock, that formally gives very fast
acceleration to extremely high energies. Leaving aside the important
question concerning the justification of the assumption for such
shock-produced turbulent motions also significantly ahead of the forward
SNR shock itself (see e.g. Ellison et al. 1994, Lucek \& Bell 2000), one
has to stress that
the final result should in any case depend upon the values of the
microscopic CR diffusion tensor which alone allows CRs to intersect the
shock front many times with a subsequent increase of their energy. As far
as the macroscopic CR diffusion due to chaotic gas motions (Biermann 1993)
is concerned, it can not itself provide CR acceleration. In fact, from
what has been said before, even our results presented below, based on the
usual assumption about Bohm type microscopic particle diffusion in a
quasi-parallel shock and much less optimistic compared with those in
Biermann (1993), must be considered as an upper limit for the CR energy
density which can be achieved during shock acceleration. This leaves open
the question whether the evolving SNR shock system can generate a strong
scattering wave field and allows the assumption of a favorable ratio of
the microscopic parallel and perpendicular diffusion coefficients over a
sufficiently long time, so that the maximum individual particle energy can
be increased by something like an order of magnitude or more for a
perpendicular shock (Jokipii 1987; Ostrowski 1988; Ellison et al. 1995; 
Reynolds 1998).

As mentioned above, a strong wind modifies the environment of massive
progenitors of SN types Ib and II. It sweeps up the ambient gas into a
thin shell surrounding a rarefied bubble (Weaver et al. 1977; Losinskaya
1991). Since the typical cavity is greater than $10$~pc in size and
contains a considerable amount of matter, it can significantly influence
the SNR evolution and CR acceleration. Estimates and preliminary numerical
calculations indicate that the SN shock propagation through the
progenitor's wind region should generate high-energy $\pi^0$-decay
$\gamma$-ray emission on a detectable level (Berezinsky \& Ptuskin 1988;
Kirk et al. 1995; Berezhko \& V\"olk 1995\footnote{Due to an error in the
plot routine in Fig. 1 of Berezhko \& V\"olk (1995), the \gr fluxes in the
wind SNR cases were plotted with an enlargement factor of 2500 which
should be left out}). Yet, up to now, all {\it detailed} numerical
investigations of CR acceleration in SNRs have dealt with the case of a
star exploding into a uniform homogeneous medium.

In this paper we present a detailed extension of the kinetic model for CR
acceleration in SNRs (Berezhko et al. 1994, 1995, 1996; Berezhko \& V\"olk
1997) to the case of a
nonuniform circumstellar medium, spherical symmetry
still being assumed. We study the CR acceleration and $\pi^0$-decay
$\gamma$-ray production taking into account that the circumstellar medium
can be strongly modified by a wind from the progenitor star. In doing this
we assume energetic particle scattering governed by the Bohm diffusion
coefficient in the local mean magnetic field, treating the shock normal to
be quasi-parallel to the field direction. We have studied two cases
typical for type Ib and type II SNe. In the first case, a star with an
initial mass of $35~M_{\odot}$ at the end of its evolution explodes as SN
into the cavity created by a main-sequence (MS) O-star wind and the
subsequent winds during the red supergiant (RSG) and Wolf-Rayet (WR)
phases. For the case of a SN type II we take the example of the explosion
of a star with initial mass $15~M_{\odot}$ into the cavity created by the
winds emitted during the MS and RSG phases. In both cases we consider the
two different interstellar number densities 0.3 and 30 cm$^{-3}$.

We shall not address here the question of $\gamma$-ray production by
electron synchrotron emission, Bremsstrahlung, or the inverse Compton (IC)
effect on ambient low energy photons. Especially at very high energies
exceeding $\sim 10$~GeV, IC $\gamma$-rays appear to dominate the emission
from plerions like the Crab nebula (e.g. de Jager \& Harding 1992, Atoyan
\& Aharonian 1996), where presumably a relativistic wind of
electron-positron pairs from the pulsar is dissipated in a circumstellar
termination shock deep inside the SNR shell. The IC effect associated with
acceleration of ultrarelativistic electrons inside the SNR shell, or at
its leading shock (along with ions considered here), may also contribute
quantitatively to the $\gamma$-ray luminosity of a SNR as simple estimates
suggest (Mastichiadis 1996; Mastichiadis \& de Jager 1996). Therefore in
any specific source this leptonic contribution to the $\gamma$-ray flux
needs to be estimated before the hadronic $\pi^0$-decay $\gamma$-ray
emission can be compared with the theoretical models presented below.

We briefly describe some aspects of the model in Sect. 2 since it
was in detail described in a previous paper (Berezhko \& V\"olk 1997).
Sect. 3 contains the results and Sect. 4 includes the discussion and
the conclusions.

\section{Model}
\subsection{CR acceleration and SNR evolution}

During the early phase of SNR evolution the hydrodynamical SN explosion
energy $E_{sn}$ is kinetic energy of the expanding shell of ejected mass.
The motion of these ejecta produces a strong shock wave in the background
medium, whose size $R_s$ increases with velocity $V_s=dR_s/dt$. Diffusive
propagation of energetic particles in the collisionless scattering medium
allows them to traverse the shock front many times. Each two subsequent
shock crossings increase the particle energy. In plane geometry this
diffusive shock acceleration process (Krymsky 1977; Axford et al. 1977;
Bell 1978; Blandford \& Ostriker 1978) creates a power law-type CR
momentum spectrum. Due to their large energy content the CRs can
dynamically modify the shock structure.

The description of CR acceleration by a spherical SNR shock wave
is based on the diffusive transport equation for the CR
distribution function $f(r,p,t)$ (Krymsky 1964; Parker 1965):
\begin{equation}
{\partial f \over \partial t}={1\over r^2}{\partial\over\partial r}
r^2 \kappa {\partial f\over\partial r}-w_c{\partial f\over\partial r}+
{1\over r^2}{\partial\over\partial r}(r^2w_c){p\over 3}
{\partial f\over\partial p}+Q,
\end{equation}
where $Q$
is the source term due to injection; $r$, $t$ and $p$ denote the radial coordinate, the
time, and particle momentum, respectively; $\kappa $ is the CR
diffusion coefficient; $u=V_s-w$. In addition,
\[
w_c=w \mbox{  for  }r<R_s,\hspace{0.3cm} w_c=w+c_a\mbox{  for  }r>R_s,
\]
where $w$ is the radial mechanical velocity of the scattering medium (i.e. thermal gas), $c_a$ is the speed of forward Alfv\'en waves generated in the upstream region by the anisotropy of the accelerating CRs. In the downstream plasma the propagation directions of the scattering waves are assumed to be isotropized (e.g. Drury et al. 1989).

The thermal matter is
described by the gas dynamic
equations
\begin{equation}
{\partial \rho \over \partial t}+{1\over r^2}{\partial\over\partial r}
(r^2 \rho w)=0,
\end{equation}
\begin{equation}
\rho{\partial w \over \partial t}+\rho w{\partial w\over\partial r} =
-{\partial\over\partial r}(P_{\rm g}+P_{\rm c}),
\end{equation}
\begin{equation}
{\partial P_{\rm g} \over \partial t}+w{\partial P_{\rm g}
\over\partial r}+{\gamma_{\rm g}\over r^2}{\partial\over\partial r}
(r^2w)P_{\rm g}=\alpha_a(\gamma_{\rm g}-1)
c_{\rm a}{\partial P_{\rm c}\over \partial r},
\end{equation}
where $\rho$, $\gamma_g$ and $P_g$ denote the mass density, specific heat ratio and the pressure of gas, respectively, and 
\begin{equation}
P_c=\frac{4\pi c}{3}
\int_0^{\infty}dp \frac{p^4f}{\sqrt{p^2+m^2c^2}}
\end{equation}
is the CR pressure. These gas dynamic equations include the CR
backreaction via term $-\partial P_c/\partial r$. They also describe the
gas heating due to the dissipation of Alfv\'en waves in the upstream
region (McKenzie \& V\"olk 1982; V\"olk et al. 1984); it is
given by the parameter $\alpha=1$ at $r>R_s$ and $\alpha_a=0$ at $r<R_s$.

We expect that the SNR shock always includes a sufficiently strong
subshock which heats the gas and plays an important dynamical role, also
in the present case of nonuniform background medium.
The gas subshock, situated at $r=R_s$, is treated as a discontinuity on
which all hydrodynamical quantities undergo a jump. 

We assume, that the injection of some (small) fraction of gas particles into the
acceleration process takes place at the subshock, that is described by the
source
\[
Q=Q_s\delta (r-R_s).
\]

 For the sake of simplicity we restrict our consideration to protons,
which are the dominant ions in the cosmic plasma. At present we only have
some experimental (e.g. Lee 1982; Trattner et al. 1994)
and theoretical (Quest 1988; Trattner \& Scholer 1991; Giacalone et al.
1993; Bennett \& Ellison 1995; Malkov \& V\"olk 1995, 1996) indications as to
what value of the injection rate can be expected. We use here a simple CR
injection model, in which a small fraction $\eta$ of the incoming protons
is instantly injected at the gas subshock with a speed $\lambda>1$ times
the postshock gas sound speed $c_{s2}$ (Berezhko et al. 1990; Kang \&
Jones 1991; Berezhko et al. 1994, 1995, 1996; Berezhko \& V\"olk 1997):

\begin{equation}
Q_s=\frac{u_1 N_{inj}}{4\pi p_{inj}^3}
\delta(p-p_{inj}) ,~
N_{inj}= \eta
N_1 , ~ p_{inj}=\lambda m c_{s2},  
\end{equation} 
where
$N=\rho/m$ is the proton number density, and $m$ is
the particle (proton) mass. For simplicity, we always use $\lambda = 2$.
The subscripts 1(2) refer to the point just ahead (behind) the subshock.

We assume that the Bohm diffusion coefficient is a good approximation for
strong shocks (McKenzie \& V\"olk 1982), characterized by strong wave
generation (Bell 1978). This latter question has been studied very
recently in its nonlinear consequences, by Lucek \& Bell (2000)
numerically, and by Bell \& Lucek (2000) using an analytical model. The
conclusion is
that wave generation in very strong shocks should not only lead to Bohm
diffusion, but also to an amplification of the pre-shock magnetic field
that increases the acceleration rate, as had been speculated earlier
(V\"olk 1984). We use here the CR diffusion coefficient
\begin{equation} 
\kappa (p)= \rho_B c/3, 
\end{equation} 
where $\rho _B$ is the gyroradius of a particle with momentum $p$ in the
magnetic field $B$, $c$ is the speed of light. This coefficient differs
from the Bohm diffusion coefficient in the non-relativistic energy
region, but this difference is absolutely unimportant because of the very
high acceleration rate at $p<mc$. In the disturbed region we use $\kappa
= \kappa_s \rho_s/\rho$, where the subscript $s$ corresponds to the
current shock position $r=R_s$. An additional factor $\rho_s/\rho$ was
assumed to prevent the instability of the precursor (Drury 1984; Berezhko
1986b). It can also be interpreted as describing the enhancement of
magnetic turbulence in a region of higher gas density.

Alfv\'en wave dissipation (V\"olk et al. 1984) as an additional
heating mechanism strongly influences the structure of a modified shock in
the case of large sonic Mach number $M =V_s/c_s \gg \sqrt {M_a}$,
$M_a=V_s/c_a$ is the Alfv\'enic Mach number, $c_s$ and $c_a$ are the local
sound and Alfv\'en speeds correspondingly, at the shock front position
$r=R_s$. The wave damping substantially restricts the growth of the shock
compression ratio $\sigma =\rho _2/\rho _s$ at the level $\sigma\approx
M_a^{3/8}$ which, in the absence of Alfv\'en wave dissipation, has been
found to reach extremely high values $\sigma \approx M^{3/4}$ for large
Mach numbers (Berezhko et al. 1996; Berezhko \& Ellison 1999).

The dynamic equations are solved under the initial ($t=t_i$)
conditions:
\[
f(p)=0,\hspace{1cm}\rho=\rho_0(r,t_w),
\]
\begin{equation}
 P_g=P_{g0}(r,t_w),\hspace{1cm}w=w_0(r,t_w),
\end{equation}
which neglect the background CRs and describe some ambient gas distribution modified by the wind from the progenitor star emitted during a preceding time
 period $t_w$. The time $t=0$ corresponds to the instant of SN explosion.

The result of a core collapse supernova, many days after the explosion, is
freely expanding gas with velocity $v=r/t$. The density profile of the
ejecta is described by (Jones et al. 1981; Chevalier 1982; Chevalier \&
Liang 1989)
\begin{equation}
\rho_{ej} =\left\{ \begin{array}{ll}
Ft^{-3}, & v<v_t\\
Ft^{-3}(v/v_t)^{-k}, & v\geq v_t,
\end{array}
\right. 
\end{equation}
where
\[
F=\frac{1}{4\pi k} \frac{[3(k-3)M_{ej}]^{5/2}} {[10(k-5)E_{sn}]^{3/2}},
~~v_t=\left[\frac{10(k-5)E_{sn}} {3(k-3)M_{ej}}\right]^{1/2},
\]
$M_{ej}$ is the total ejected mass. For SNRs the value of the parameter
$k$ typically lies between 7 and 12. The pressure in the expanding
ejecta is negligible.

Interaction with the ambient material modifies the ejecta density
distribution. We describe the ejecta dynamics in a simplified manner,
assuming that the modified ejecta consist of two parts (Berezhko \& V\"olk
1997): a thin shell (or piston) moving with some speed $V_p$ and a freely
expanding part which is described by the distribution (9). The piston
includes the decelerated tail of the distribution (9) with initial
velocities $v>R_p/t$, where $R_p$ is the piston radius separating the
ejecta and the swept-up ISM matter. The evolution of the piston is
described in the framework of a simplified thin-shell approximation, in
which the thickness of the shell is neglected. Behind the piston
$(r<R_p)$, the CR distribution is assumed to be uniform.

The high velocity tail in the distribution (9) ensures a large value of
the SNR shock speed at an early phase of evolution. It increases the CR
and $\gamma$-ray production significantly compared with the case where all
the ejecta propagate with a single velocity (Berezhko \& V\"olk 1997).

In the case of a uniform ISM it was shown that the CR penetration through
the piston plays no important role for SN shock evolution and overall CR
production (Berezhko et al. 1996). In the case of an intensive wind from
the progenitor star, a relatively large volume around the progenitor is
occupied by a low density bubble. The main amount of CRs and $\gamma$-rays
are produced when the SNR shock interacts with this bubble. Piston and
shock sizes are comparable and much larger than the dynamic scale of the
system during the most effective CR production phase. Therefore one can
expect that CR penetration through the piston is more important in
comparison with the case of a uniform ISM.

The efficiency of diffusive CR penetration through the piston depends on
the magnetic field structure, which is influenced by the
Rayleigh-Taylor instability at the contact discontinuity ($r=R_p$) between
the ejecta and the swept up medium that is contained in the region
$R_p<r<R_s$.  According to Chevalier's (1982) estimate, the energy density
of the turbulent motions created by this instability is determined by the
value of the thermal pressure $P_p$ at the outer piston surface
$(r=R_p+0)$,
and may be as large as $e_t=0.2P_p$.  Turbulent motions in the ionized
medium lead to the amplification of magnetic fields. We assume that the
magnetic field grows up to the energy density $e_B=0.5e_t$. This turbulent
magnetic field is presumably distributed over a wide range of length
scales. The energy density of the magnetic field fluctuations which
resonantly
interact with particles in the momentum range from $mc$ to $p_m$ may be
as large as $e_B/\ln (p_m/mc)$, which is about $0.1e_B$ for a typical CR
cutoff momentum $p_m=10^4mc$.  Therefore we use a Bohm type diffusion
coefficient in the piston region, corresponding to a magnetic field
strength of $B=\sqrt{8\pi \delta P_p}$ with $\delta=10^{-2}$.

Radiative gas cooling is not included in our model. This process becomes
important at the late Sedov phase (Dorfi 1991), when CR acceleration becomes
inefficient.

Detailed descriptions of the model and of the numerical methods have been
given earlier (Berezhko et al. 1994, 1995, 1996; Berezhko \& V\"olk 1997).

\subsection{Progenitor wind bubble} 

The strong wind from the massive progenitor star interacts with an ambient
ISM of uniform density $\rho_0 = 1.4mN_H$, resulting to first
approximation in an expanding spherical configuration, which is called a
bubble (Weaver et al. 1977; Losinskaya 1991). Here $m$ is the proton mass
and $N_H$ is the hydrogen number density in the background ISM where 10\%
of helium by number is assumed. Throughout its evolution, the system
consists of four distinct zones.  Starting from the center they are:  (a)
the hypersonic stellar wind (b) a region of shocked stellar wind (c) a
shell of shocked interstellar gas, and (d) the ambient ISM.

We consider here the case of a so-called modified bubble whose structure
is significantly influenced by mass transport from the dense and
relatively cold shell (c) into the hot region (b), and by thermal
conduction in the opposite direction, these two energy fluxes balancing
each other to first order. During this stage the shell (c) has collapsed
into a thin isobaric shell due to radiative cooling (Weaver et al. 1977;
Kahn \& Breitschwerdt 1989).

The sizes of each zone are determined by the ISM number density $N_H$, the wind
speed $V_w$ and the stellar mass-loss rate $\dot{M}$ (Weaver et al. 1977),
and
can be written in the form:

 \[ R_1=4.38\left(\frac{\dot{M}}{10^{-6}M_{\odot}/\mbox{yr}}\right)^{3/10}
 \times
 \]
 \begin{equation}
\left(\frac{V_w}{2000\mbox{ km/s}}\right)^{1/10}
\left(\frac{N_H}{1\mbox{ cm}^{-3}}\right)^{-3/10}
\left(\frac{t_w}{10^6\mbox{ yr}}\right)^{2/5}\mbox{ pc}
\end{equation}
is the inner shock radius which bounds the wind region (a);
\[
R_2=27.5\left(\frac{\dot{M}}{10^{-6}M_{\odot}/\mbox{yr}}\right)^{1/10}
 \times
 \]
\begin{equation}
\left(\frac{V_w}{2000 \mbox{ km/s}}\right)^{4/10}
\left(\frac{N_H}{1\mbox{ cm}^{-3}}\right)^{-1/10}
\left(\frac{t_w}{10^6\mbox{ yr}}\right)^{3/5} \mbox{ pc}
\end{equation}
is the outer shock radius which separates the shell (c) from the
 ISM.
The time $t_w=0$ corresponds to the onset of a steady,
spherically symmetric stellar wind.

 Region (a) is characterized by a negligibly small gas pressure
$P_g$,
a constant speed $V_w$, the gas density
\begin{equation}
\rho=\frac{\dot{M}}{4\pi V_w r^2}
\end{equation}
and an essentially toroidal magnetic field (we use its value near the
equatorial plane)
\begin{equation}
B=B_\ast \frac{R_\ast \Omega}{V_w}\frac{R_\ast}{r},
\end{equation}
where $R_\ast$ is the radius and $\Omega$ the angular
rotation rate of the star.

The regions (b) and (c) are isobaric with a thermal pressure (Weaver et
al. 1977) that can be written in the form
\begin{eqnarray} 
P_b & = & 4.7\times
10^{-12}\left(\frac{\dot{M}}{10^{-6}M_{\odot}/\mbox{yr}}\right)^{2/5}
\left(\frac{V_w}{2000\mbox{ km/s}}\right)^{4/5}\nonumber \\
   &  \times  &
\left(\frac{N_H}{1\mbox{ cm}^{-3}}\right)^{3/5}
\left(\frac{t_w}{10^6\mbox{ yr}}\right)^{-4/5}\frac{\mbox{ dyne}}{\mbox{cm}^2}.
\end{eqnarray}
The gas number density of the region (b) $N_b=\rho_b/m$ is
approximately uniform and given by
\begin{eqnarray}
N_b & = & 3.8\times 10^{-2}\left(\frac{\dot{M}}{10^{-6}M_{\odot}/\mbox{yr}}
\right)^{6/35}
\left(\frac{N_H}{1\mbox{ cm}^{-3}}\right)^{19/35}\nonumber \\
  &  \times &
\left(\frac{V_w}{2000\mbox{ km/s}}\right)^{12/35}
\left(\frac{t_w}{10^6\mbox{ yr}}\right)^{-22/35}\mbox{ cm}^{-3}.
\end{eqnarray}
The shell (c) is much denser: $N_c\gg N_b$. Therefore we can describe
the number density distribution in regions (b) and (c) by the combined
expression
\begin{equation}
N_g=\sigma_c N_0 \left( \frac{r}{R_2} \right) ^{3(\sigma_c -1)}+N_b,
\end{equation}
where $N_0=1.4N_H$, $\sigma_c N_0$ is the peak value of the shell number
density reached at the front of the outer shock $(r=R_2-0)$, and 
$\sigma_c$ is the
outer shock compression ratio.

The mass and heat transport between regions (b) and (c) are presumably
due to turbulent motions in the bubble.  We assume that
the turbulent motions generate a magnetic field which grows up to the
equipartition value
\begin{equation} 
B_b=\sqrt{8\pi P_b}.  
\end{equation}

\subsection{Gamma-ray production}
$\pi^0$-decay gamma rays are produced by energetic CR protons in inelastic
collisions
with gas nuclei which generate also neutral pions that subsequently
decay.
The $\gamma$-ray emissivity of a SNR (in units of photons/s) can be
written as (Drury et al. 1994)
\begin{equation}
Q_{\gamma}( \epsilon _{\gamma})=16\pi ^2 \int_{0}^{\infty} dr r^2
\int_{p_{\gamma}}^{\infty} dp p^2\sigma_{pp} Z_{\gamma}^{\alpha}
c N_g f(r,p,t),
\end{equation}
where
\begin{equation}
\sigma_{pp}=38.5+0.46\ln^2 (0.01876 p/ {\rm mc})~~\mbox{mb}
\end{equation}
is the inelastic $p-p$ cross-section (e.g. Berezinsky et al. 1990),
$Z_{\gamma}^{\alpha}$ is the so-called spectrum-weighted moment of the
inclusive
cross-section, $p_{\gamma}$ is the momentum of a CR particle with
kinetic energy $\epsilon= \epsilon_{\gamma}$, $N_g=\rho /{\rm m}$ is the
gas number density, $ \alpha=1-d\ln n / d\ln \epsilon $ is the power
law index of the integral CR energy spectrum, and $n=4\pi p^2 f$ denotes 
the CR
differential number density.  We use the approximation for the 
spectrum-weighted moment as given by Drury et al. (1994), limiting
$Z_{\gamma}^{\alpha}$ by the value 
\begin{equation}
Z_{\gamma}^{\alpha}=min\{0.2,10^{1.49-2.73\alpha+0.53\alpha^2}\}
\end{equation}

The integral $\gamma$-ray flux at the distance $d$ from the source is
\begin{equation}
F_{\gamma}(\epsilon_{\gamma})=Q_{\gamma}(\epsilon_{\gamma})/4\pi d^2.
\end{equation}

\section{Results} 

Detailed investigations of CR acceleration and SNR
evolution in a uniform ISM have revealed important features of this
process (Berezhko et al. 1994, 1995, 1996; Berezhko \& V\"olk 1997). For a
wide range of injection rates the CR acceleration efficiency is very high,
near unity, and therefore almost independent of the injection rate.
Therefore we use here the particular value of the injection parameter
$\eta=10^{-3}$, which corresponds to a moderate injection rate for a
parallel shock.  This value of $\eta$ implies an injection rate which is
almost one order of magnitude lower than that resulting from simulations
of collisionless plasma shocks (Quest 1988; Trattner \& Scholer 1991;
Giacalone et al. 1993), and which also corresponds to the kinetic MC model
(e.g. Ellison et al. 1995; Baring et al. 1999) for parallel shocks. Our
lower value of $\eta$ effectively takes into account the influence of the
average shock obliquity for a quasi-spherical SNR which expands into an
ISM with a uniform mean magnetic field: according to Ellison et al.
(1995), and
Malkov \& V\"olk (1995,) already at an angle $\theta \approx 45^o$ between
the upstream magnetic field and the shock normal the injection rate is
about an order of magnitude smaller than in the purely parallel shock
case, if particle diffusion is somewhat less efficient than the Bohm
limit. We can not exclude that even this effective injection rate leads to
an overestimate for the global acceleration efficiency of the SNR.

We also restrict our consideration to a typical set of values of the SN
parameters: hydrodynamic explosion energy $E_{sn}=10^{51}$~erg, ejecta
mass $M_{ej}=10M_{\odot}$, $k=10$.

The radiative cooling of the swept-up shell of interstellar gas leads to
high shell compression ratio $\sigma_c \gg 1$. The ISM magnetic field and
the CRs which are produced by the outer shock can nevertheless
significantly restrict this shell compression ratio. We use here the
moderate value $\sigma_c =10$. In fact, the final results are quite
insensitive to the value of $\sigma_c$ because the SN shock becomes an
inefficient CR accelerator before reaching the region $r=R_2$ of the peak
shell density.

\subsection{Type Ib supernova}
As a typical example for a type Ib SN we use theoretical results of
stellar evolution with initial mass $35M_{\odot}$ (Garcia-Segura et al.
1996). According to these calculations the evolution consists of three
stages: a MS phase with mass-loss rate $\dot{M}=5.56\times
10^{-7}M_{\odot}$~yr$^{-1}$, wind speed $V_w=2000$~km/s, and duration
$\Delta t_w=4.5\times 10^6$~yr; a RSG phase with
$\dot{M}=10^{-6}M_{\odot}$~yr$^{-1}$ $V_w=15$~km/s, $\Delta t_w=2\times
10^5$~yr, and a WR-phase with $\dot{M}=2.25\times
10^{-5}M_{\odot}$~yr$^{-1}$, $V_w=2000$~km/s, and $\Delta t_w=2\times
10^5$~yr.

According to Eq. (11) the MS wind creates a bubble of size $R_2=76.8$~pc.
The parameters of this modified bubble are $N_b=5\times
10^{-3}$~cm$^{-3}$, $P_b=5.47\times 10^{-13}$~dyne/cm$^2$.

The wind emitted during the RSG phase occupies a region of size
$R_f=\Delta t_w V_w=0.3$~pc.

The fast wind from the subsequent WR star interacts with this dense RSG
wind. After a relatively short period of time the WR wind breaks through
the RSG wind material, leaving it in the form of clouds (Garcia-Segura et
al. 1996). Taking into account that the interaction time is short and that
the mass of the RSG wind is small compared with the bubble mass, we
neglect in a zeroth approximation the influence of RSG phase on the final
structure of the bubble.

Using again approximately Eqs. (11) and (15), the WR wind inside the
MS-bubble now creates a WR bubble with number density $N_b=1.02\times
10^{-2}$~cm$^{-3}$ and a shell of size $R_2=56.2$~pc. Formally this value
$N_b$ is twice as large as the number density of the MS bubble. If we
disregard in our approximation the stellar mass lost in the WR phase
compared to the mass of the partially swept-up MS bubble, the density of
the new WR bubble can not exceed that of the MS bubble shell, simply for
mass conservation. It can at best be equal if we assume that the WR shell
is completely dissipated, i.e. smeared out over the WR bubble. This is in
fact a quite reasonable assumption and we shall adopt it here. Therefore
we use a simplified structure of the bubble, whose parameters
$R_2=76.8$~pc, $N_b=5\times 10^{-3}$~cm$^{-3}$, $P_b=5.47\times
10^{-13}$~dyne/cm$^2$, $B=7.89$~$\mu$G correspond to the MS-bubble,
together with a hypersonic WR wind region of size $R_1=30.8$~pc,
corresponding to the pressure equilibrium condition $\rho V_w^2=P_b$. This
approximation to the dynamics is justified to the extent that we can
neglect the overall mass lost by the star in comparison with the overall
swept-up interstellar mass.

The assumed circumstellar density distribution $N_g$ is shown on Fig. 1
as a function of radial distance $r$ for all the cases considered.

\begin{figure}
 \resizebox{\hsize}{!}{\includegraphics{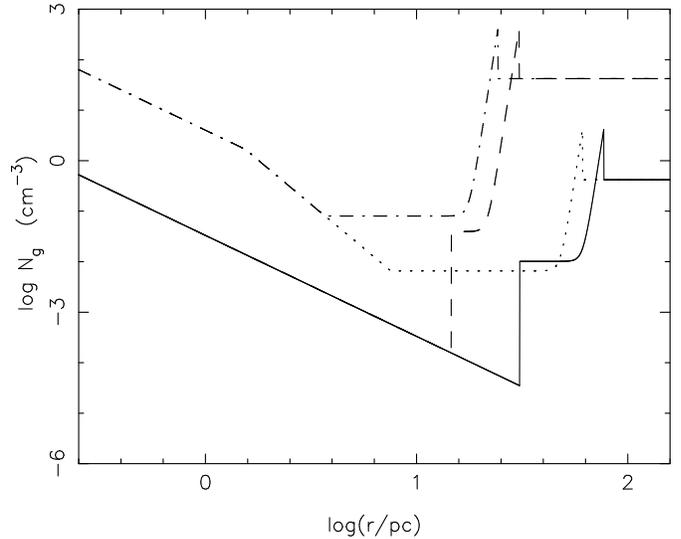}}
 \caption{Circumstellar gas density distribution as a function of radial
 distance for the case of a SNR Ib  in an ISM with hydrogen number
 densities
 $N_H=0.3$~cm$^{-3}$ (full line) and
 $N_H=30$~cm$^{-3}$ (dashed line), and for the case of a SNR II with
 $N_H=0.3$~cm$^{-3}$
 (dotted line) and $N_H=30$~cm$^{-3}$ (dash-dotted line).}
\end{figure}

For the WR-star we use the parameters $R_*=3\times 10^{12}$~cm, $\Omega
=10^{-6}$, $B_*=50$~G, which determine the value of magnetic field in
the wind region (a).

As we have already mentioned in the Introduction it is not obvious that
our model (which is strictly speaking only valid for the case of a
quasi-parallel shock) can be applied to the present case of a shock that
is on average almost purely perpendicular. Basically it is not clear
whether a sufficient number of suprathermal particles can be injected into
a quasi- perpendicular shock (Ellison et al. 1995; Malkov \& V\"olk 1995).
On the other hand, there exist circumstances which should alleviate the
obstacles for particle injection and their subsequent acceleration. One
may assume that the magnetic field is essentially disordered on a spatial
scale $l_B$, small compared with the main scale $r$. Over a substantial
fraction of the magnetic flux tubes the SN shock will then be locally
quasi-parallel, with particle injection and acceleration starting there
without difficulties. If the scale $l_B$ is larger than the diffusive
length of suprathermal particles $l(p_{inj})=\kappa_s(p_{inj})/V_s$, these
particles will be at least accelerated up to the momentum $p_*$, where
$l(p_*)\sim l_B$.  If $l_B\gg l(p_{inj})$, accelerated particles with
momenta $p_{inj}\leq p \leq p_*$ will gain a substantial part of the shock
energy. In this case they can amplify magnetic field disturbances on
progressively larger scales, that will allow another fraction of these
particles to be further accelerated. Ultimately, some particles might
indeed reach the maximum energy, which is determined by the usual physical
factors for a quasi-parallel shock (see below). However the particle
spectrum is expected to be softer than for a quasi-parallel shock. For the
suprathermal particles with energy $p_{inj}\sim mV_s$ the diffusive length
is about $l\sim 10^{-6}R_s$. Therefore the above scenario will take place,
if the magnetic field is initially disordered on a scale $l_B\gsim
10^{-4}r$, that seems not an unreasonable assumption.

We note, that there is some experimental evidence that particle (at least
electron) diffusive shock acceleration in the progenitor wind actually
takes place (see e.g. Chevalier 1982, Kirk et al. 1995, and references
there).


We shall use here the assumption of Bohm type CR diffusion for a locally
quasi-parallel shock. It implies an acceleration rate that is almost
independent of the magnetic field structure. However, in the light of our
above discussion of this field structure we consider our results as upper
limits to the real CR and gamma-ray production efficiency.

Starting from some initial instant $t_i=R_{pi}/V_{pi}=6.34$~yr after the
SN explosion which corresponds to an initial piston position
$R_{pi}=3\times10^{17}$~cm and a piston speed $V_{pi}=15\times10^3$~km/s,
we explore the solution of the nonlinear dynamic equations following the
SN shock propagation through the successive bubble zones. Note that the
value of the initial position $R_{pi}$ from which we start our
calculations has as little physical importance as the short subsequent
period of SNR evolution during which the quasi-stationary character of
system evolution is established.

Fig. 2 illustrates the CR characteristics, SN shock parameters and
expected
$\gamma$-ray spectra produced during SN shock propagation through the
supersonic WR wind region $r<R_1$.
\begin{figure*} 
\resizebox{\hsize}{!}{\includegraphics{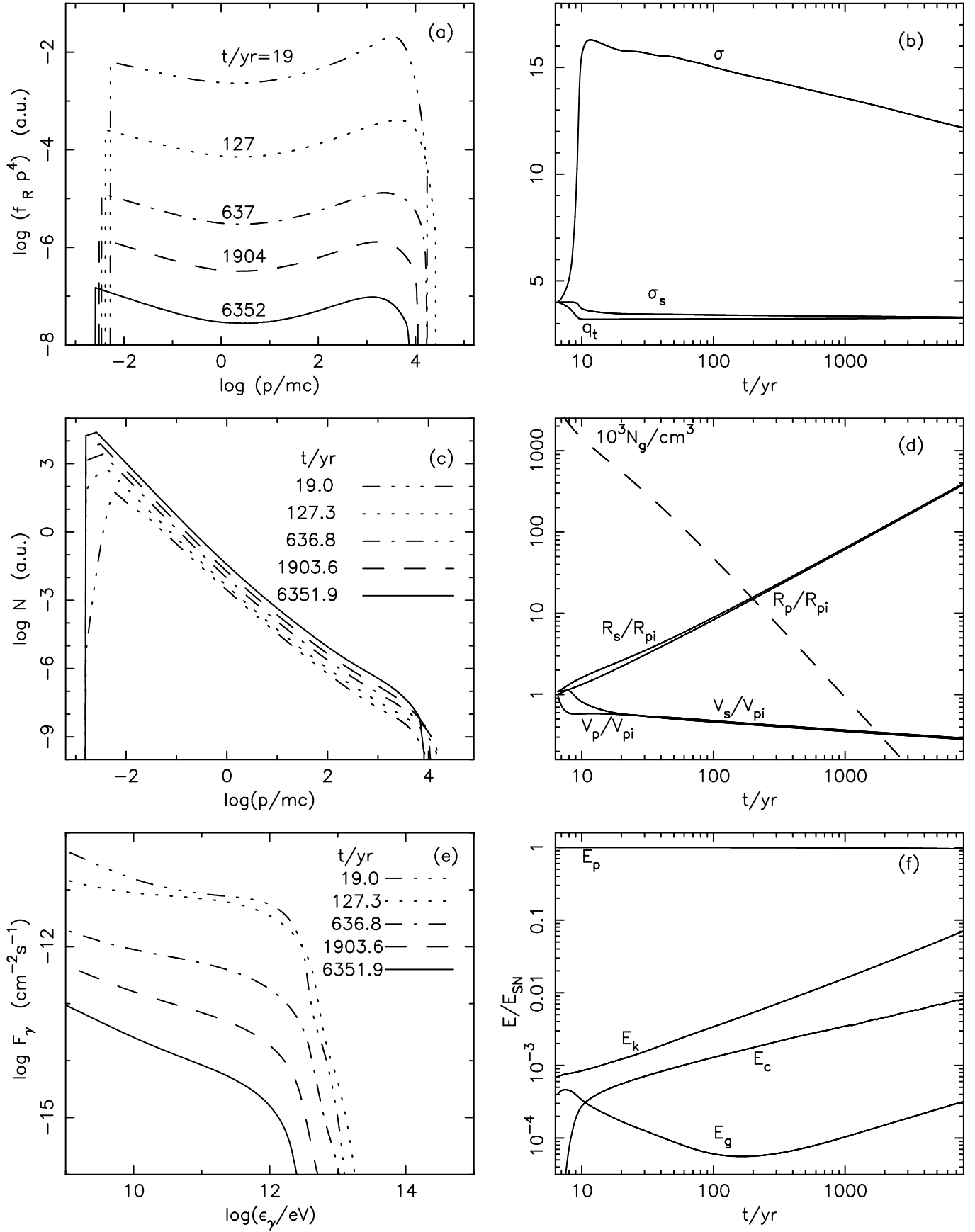}}
\caption{CR and gas dynamics in a WR wind region. CR distribution function
at the shock front (a), volume integrated overall CR spectrum (c), as
functions of momentum, and overall integral $\gamma$-ray spectrum
normalized to a distance of 1 kpc (e), as function of $\gamma$-ray energy,
at five different times $t$. Total shock compression ratio $\sigma$,
subshock compression ratio $\sigma_s$, and differential power law index
$q_t$ (b). Shock (suffix $s$) and piston (suffix $p$) radii $R$ and
velocities $V$ (full lines) and circumstellar gas number density $N_g$
(dashed line) (d). Ejecta energy $E_p$, kinetic $(E_k)$ and thermal
$(E_g)$ gas energies, and CR energy $E_c$ (f) as functions of time, for a
SN shock propagating through the supersonic wind region of a WR progenitor
star with $\dot{M} =10^{-6}~M_{\odot}$~yr$^{-1}$,
$\Omega=10^{-6}$~s$^{-1}$, $R_*=3\times 10^{12}$~cm, $B_*=50$~G, and
$V_w=2000$~km/s. }
\end{figure*}

In Fig. 2a the calculated CR distribution function at the subshock
$f_s(p,t)=f(r=R_s,p,t)$ is presented for five different instants of time.
The shape of the CR spectrum is determined by the fact that the shock is
essentially modified by CR backreaction. Very soon after the beginning of
CR acceleration the total shock compression ratio $\sigma=\rho_2/\rho_s$
reaches about $\sigma =16.5$ and then slowly decreases with time. Although
the total compression ratio is much higher than predicted by the
hydrodynamical two-fluid model (Jones \& Kang 1992), the subshock never
disappears and its compression ratio remains nearly constant at the level
$\sigma _s=\rho_2/\rho_1 \approx3.4$ (see Fig. 2b).  At small momenta
$p_{inj}\leq p \lsim mc$ the CR distribution function is almost a pure
power-law $f_s\propto p^{-q_s}$ with the power-law index 
\begin{equation} q'_s=3\sigma'_s/(\sigma'_s-1), \end{equation} 
determined by the effective subshock compression ratio $\sigma'_s=\sigma_s
(1-1/M_{a1})$. At momenta $10~{\rm mc}\lsim
p\lsim p_m$, where $p_m\sim10^4{\rm mc}$, the distribution function is
also close to a power-law form $f_s\propto p^{-q}$ with the index value
$q=3.6$.

It is important to note, that our results confirm the conclusion that the
accelerated particle spectrum can not be harder than $f\propto p^{-3.5}$
even for a very large shock compression ratio $\sigma$ (Berezhko 1996;
Malkov 1997; Berezhko \& Ellison 1999a), whereas the test particle
approach (see expression (27) below) in the case $\sigma \gg 1$ predicts
the limiting value of the power law index $q=3$.

An interesting point is that in the wind the CR distribution functions
$f_s(p)$ at different evolutionary phases are almost selfsimilar to each
other except during the initial short period $t<10$~yr when the system
undergoes the transition from the initial conditions at $t=t_i$ to the
quasi-stationary state.  It is a result of the fact, that during the
propagation through the low density supersonic wind the shock speed
remains nearly constant. The analysis of the dynamical equations shows
that in this case the CR distribution function has a selfsimilar form
$f(r,p,t)=\Phi(r/R_s,p)/t^2$ (see Appendix) which is roughly consistent
with the numerical results presented on Fig. 2a. It also explains the time
constancy of the maximum particle momentum obtained by V\"olk \& Biermann
(1988), as mentioned in the Introduction..

A more detailed consideration includes the shock deceleration. In the
case of the ejecta density described by the selfsimilarity
distribution  Eq. (9) at
the early phase, when $V_s>v_t$, the expected expansion law is
\begin{equation}
R_s\propto \left(\frac{V_w}{\dot{M}}\right)^{1-\nu }t^{\nu}
\end{equation}
with $\nu =(k-3)/(k-2)$ (Chevalier 1982), which in our case has the value 
$\nu =0.875$. The decrease
of the shock velocity 
\begin{equation} V_s=\nu \frac{R_s}{t}\propto
t^{\nu -1} 
\end{equation} 
leads to a decrease of the injection momentum $p_{inj} \sim mV_s\propto
t^{\nu -1}$ that introduces an additional time dependent factor in the CR
distribution function $f\propto p_{inj}^{q_s-3}\propto t^{(\nu
-1)(q_s-3)}$ at least at nonrelativistic energies. It results in the
dependence $f\propto t^{-\mu }$ where $\mu = 2\nu -(\nu -1)(q_s-3)$. In
our case $k=10$, $\sigma _s =3.4$ this gives $q_s=4.25$ and $\mu = 1.91$.
Due to the selfregulating property of the nonlinear acceleration process,
CR particles absorb some constant fraction of the shock energy; therefore
at relativistic energies $f_s\propto \rho_s V_s^2 \propto t^{-2}$.

A direct consequence of this fact is the evolution of the overall, i.e.
volume integrated CR spectrum 

\[ N(p,t)=16\pi^2 p^2 \int_0^{\infty}dr r^2f(r,p,t), 
\] 

which is shown in Fig. 2c as a function of momentum $p$ for five
subsequent evolutionary phases. The shape of the spectrum and the value of
the cutoff momentum $p_m$ change slowly during the evolution. At high
energies\\ $10^2\lsim p/{\rm mc} \lsim 10^4$ the overall CR spectrum is
close to the form $N\propto p^{-1.5}$. It is interesting to note that the
overall CR spectrum is even harder than the local CR spectrum at the shock
front $n_s= 4\pi p^2 f_s(p)$, because the size of the upstream region
$l=\kappa_s(p)/V_s$ occupied by CRs with momentum $p $ is proportional to
$p$. At relatively low momenta $l(p)<<R_s-R_p$ and the overall spectrum
$N(p)$ mainly consists of the CRs situated in the downstream region
$R_p<r<R_s$; therefore in this case $N(p)\propto n_s(p)$. At high momenta
$p\sim p_m$ diffusive length $l(p)$ becomes comparable with the downstream
region size $R_s-R_p\approx R_s/(3\sigma)$; therefore the overall spectrum
$N(p)$ becomes progressively harder with increasing $p$ due to upstream
region. The amplitude of the spectrum grows with time because the volume
occupied by accelerated CRs increases with time $V\propto R_s^3$ that
gives $N\propto Vf_s p^2\propto t^{3\nu -\mu}$.  The index $3\nu -\mu$
varies from 0.72 at $p<{\rm mc}$ to 0.63 at $p>{\rm mc}$ in good agreement
with the numerical results.

One can compare the calculated and the expected value of the CR cutoff
momentum, which is determined by the expression (Berezhko 1996) 
\begin{equation}
\frac{p_m}{mc}=\frac{R_s(V_s-V_w)}{A\kappa_s(mc)},
\end{equation} where
\begin{equation}
A=\left[ 2+2b+e+d-(\nu -1)/\nu \right]q_t/(5-q_t),
\end{equation}
and the parameter $d=(r\nabla \vec {w})_2\sigma /[(\sigma-1)V_s]$
describes
the effect of particle adiabatic cooling in the downstream region, and the
dimensionless parameters \[ \nu=d\ln R_s/d\ln t,~b=d\ln f_s/d\ln
R_s,~e=\nu d\ln \kappa_s/d\ln R_s \] describe the time variation of the
shock radius, of the CR distribution function and of the CR diffusion
coefficient, respectively. In addition
\begin{equation}
q_t=3\sigma'/(\sigma'-1)
\end{equation}
is the lower limit for the power law index, and
$\sigma'=(V_s-V_w-c_a)/u_2=\sigma (1-1/M_a)$ is the effective total shock
compression ratio which includes the effect produced by the outward
propagation of Alfv\'en waves in the upstream region $r>R_s$. By definition
the value of $p_m$ is that momentum where the local power law index
$q=-d(\ln f)/d(\ln p)$ drops to the value the value $q=5$.

In the case under consideration the gas velocity $w$ is almost constant in
the downstream region, which gives $d=2$ (Berezhko 1996). Taking into
account that $R_s\propto t^{0.875}$, $f_s\propto t^{-2},~\kappa_s \propto
R_s$ we have $\nu=0.875,~b=-2, {\rm and}~e=0.875$, and the expression for
the expected cutoff momentum can be written in the form
\begin{eqnarray} 
\frac{p_m}{mc} & = & 2.5\times 10^4 \left( \frac{B_*}{50~\mbox{G}}\right) 
\left(\frac{\Omega}{10^{-6}~\mbox{s}^{-1}}\right) \\
            & \times &
 \left(
 \frac{R_*}{3\times10^{12}~\mbox{cm}}\right) ^2
 \left( \frac{2\times
 10^3~\mbox{km/s}}{V_w}\right) \left(
 \frac{t}{10~\mbox{yr}}\right)^{-0.125}\nonumber
 \end{eqnarray}

which is in a good agreement with the numerical results (see Fig. 2a,c).  
Note that the main factors which determine the value of $p_m$ are the
adiabatic CR cooling in the expanding medium, the finite time increase of
the shock size, and the decrease of the shock velocity, but not the time
factor, as it is frequently assumed (e.g. Lagage \& Cesarsky 1983).

During the time interval $t<5922$~yr of shock propagation
through region (a) of the supersonic wind CRs absorb only a small
fraction of the explosion energy $E_c\approx 0.008E_{sn}$ (see Fig. 2f)
due to the small amount of swept up matter $M_{sw}=0.3 M_{\odot}$
compared to the ejected mass $M_{ej}=10M_{\odot}$, where
\begin{equation} M_{sw}=4\pi \int_{0}^{R_s} drr^2\rho,
\end{equation}
$\rho (r)$ is the wind density, determined by the expression (12).

The $\gamma$-ray spectrum $F_{\gamma}$ is shown in Fig. 2e. It is
extremely
hard in the energy range $10^{10}\lsim \epsilon_{\gamma}\lsim 10^{12}$~eV
during the early phase $t<500$~yr and becomes progressively steeper as
time proceeds. This behavior is a consequence of the fact that the
$\gamma$-ray flux $F_{\gamma}=F_{\gamma}'+F_{\gamma}''$ consists of two
different parts. The first, $F_{\gamma}'$, is due to CR interaction with
the swept-up matter which lies between the piston and shock surfaces and
has the mass $M_{sw}$. In the case we consider here, the CRs are almost
uniformly distributed in the downstream region $R_p<r<R_s$. Therefore we
have approximately (Berezhko \& V\"olk 1997)
\begin{equation} 
F_{\gamma}'\propto
M_{sw}e_{c2}p_m^{\gamma -2}, 
\end{equation}
where $e_{c2}$ is the CR energy density in the downstream region. In our
case $e_{c2}\propto \rho_s V_s^2$, $M_{sw}=\dot{M}R_s/V_w$, $\gamma
\approx 1.7$, giving the expected dependence $F_{\gamma}'\propto t^{-1.1}$
which is in a agreement with the numerical results at $t>10$~yr.  The
calculated $\gamma$-ray flux can be represented in the form
\begin{equation} F_{\gamma}'(1~\mbox{TeV})=F_{10}' \times  \left(
\frac{d}{1~\mbox{kpc}}\right)^{-2}
\left(\frac{t}{10~\mbox{yr}}\right)^{-1.1}
\end{equation}
with $F_{10}'=2\times 10^{-12}$~cm$^{-2}~$s$^{-1}$,
which is in rough agreement with the estimates of Berezinskii \& Ptuskin
(1989), and Kirk et al. (1995) which do not
include CR penetration into the ejecta with subsequent $\gamma$-ray
generation. The energy spectrum of this component $F_{\gamma}'\propto
\epsilon_{\gamma}^{-0.6}$ is steeper than the CR integral spectrum,
because upstream CRs interacting with the relatively low density medium
contribute much less to the $\gamma$-ray spectrum than to the
overall CR spectrum

The second part of $\gamma$-ray flux $F_{\gamma}''$ has its origin in the
ejecta material whose CR energy density is $e_{c3}$.  Therefore we have
\begin{equation}
F_{\gamma}''\propto M_{ej}e_{c3}p_m^{\gamma-2}.
\end{equation}
\noindent The ratio of the two components
\[
F_{\gamma}''/F_{\gamma}'=(M_{ej}/M_{sw})(e_{c3}/e_{c2})
\]

\noindent is determined by the two factors $M_{ej}/M_{sw}$ and
$e_{c3}/e_{c2}$. The swept up mass $M_{sw}=\dot{M}R_s/V_w$ increases with
time and reaches the value $M_{sw}=0.03M_{ej}$ at $t\sim 6000$~yr. At
$t<100$~yr the ratio $e_{c3}/e_{c2}$ grows from zero to about 0.05 and
then remains nearly constant. Therefore, during the initial period
$t<10^3$~yr, the $\gamma$-ray flux at $\epsilon_{\gamma}\sim 1$~TeV is
dominated by the second component whereas at $t>6000$~yr the first
component becomes dominant at almost all energies.

Since $e_{c3}\propto \rho_s V_s^2\propto t^{-2}$ at $t>100$~yr we have
$F_{\gamma}''\propto t^{-2}$ and numerically
\begin{eqnarray}
F_{\gamma}''(\mbox{1 TeV}) & =  & 4.6\times 10^{-12} \nonumber \\
                        & \times  &
\left( \frac{d}{1~\mbox{kpc}}\right)^{-2}
\left(\frac{t}{100 \mbox{ yr}}\right)^{-2}
\frac{1}{\mbox{cm}^2\mbox{s}}.
\end{eqnarray}

The second component $F_{\gamma}''(\epsilon_{\gamma})$ is much harder than
$F_{\gamma}'(\epsilon_{\gamma})$ (see Fig. 2e) because CRs with higher
energy penetrate into the piston more effectively. Therefore the expected
flux $F_{\gamma}(\epsilon_{\gamma})$ becomes progressively steeper as the
contribution of $F_{\gamma}''$ decreases with time (see Fig. 2e).

The time dependence of the expected integral flux of TeV $\gamma$-rays at
the distance $d=1$~kpc is shown in Fig. 3. It also includes the epochs
when the SNR shock leaves the wind region and enters the bubble and the
shell regions, to which we turn now.

\begin{figure}
\resizebox{\hsize}{!}{\includegraphics{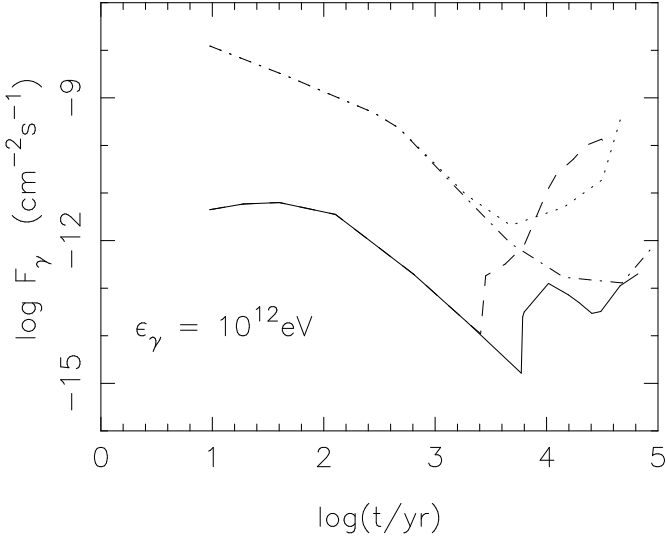}}
\caption{Integral flux of $\gamma$-rays with energy
$\epsilon_{\gamma}>$1 TeV, normalized to a distance of 1 kpc, as a
function of time for the cases of a SNR Ib with $N_H=0.3$~cm$^{-3}$
(full line), SNR Ib with $N_H=30$~cm$^{-3}$ (dashed line),
SNR II with $N_H=0.3$~cm$^{-3}$ (dash-dotted
line), and SNR II with $N_H=30$~cm$^{-3}$ (dotted line).}
\end{figure}

For $t>t_1=5922$~yr the SN shock propagates through the hot region (b) of
the MS bubble. When the SNR shock intersects the boundary $r=R_1$, which
is also a strong discontinuity, a new secondary shock arises (Shigeyama \&
Nomoto 1990). It propagates several times between the SNR shock front and
the piston surface, and compresses and heats the medium. It thus provides
the transition to the new quasi-stationary state which corresponds to the
SNR shock propagating through region (b). We neglect this complicated
transition phenomenon. We rather describe the beginning of the SN shock
propagation in the region (b) in the following simplified manner. We start
with a pure gas shock of size $R_s=R_p+0$, piston size $R_p=R_1$ and
piston speed $V_p=V_{p1}=4127$~km/s which is the final piston speed in the
region (a). We neglect the CRs produced during the previous period
$t<t_1$. This underestimates the CR and $\gamma$-ray production in region
(b). But the influence of previously produced CRs is not very large
because the CR production rate increases sharply when the SNR shock
intersects the boundary $r=R_1$, on account of the much higher gas density
in region (b) (see Fig.1).

Fig. 4 illustrates the dynamics of the system during this period.
\begin{figure*}
 \resizebox{\hsize}{!}{\includegraphics{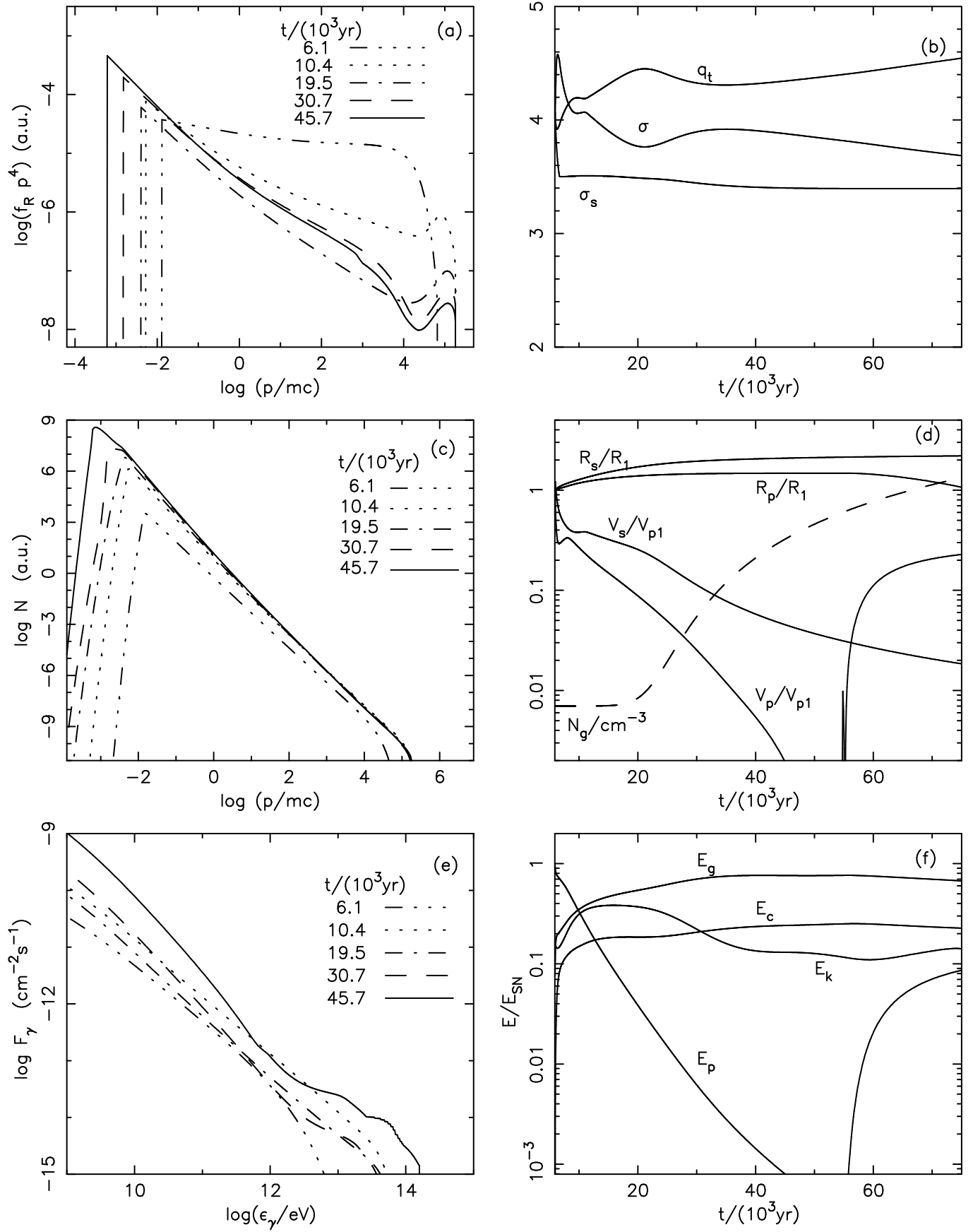}}
\caption{The same as Fig.2 but for SNR shock propagation through the WR
bubble which was reached after 5922 yr. The normalization parameters in
Fig. 4c are, respectively, $R_1=30.8$~pc, $V_{p1}=4127$~km/s.}
\end{figure*}

Note that the CR distribution functions in Figs. 2a and 4a are measured
in the same units; this is also the case for the overall CR spectra in
Figs. 2c and 4c. One can see that already at $t=6100$~yr the CR number
density and the CR distribution function increase by more than two orders
of magnitude compared to the time $t=t_1=5922$~yr. In the region (b)  
efficient CR acceleration lasts less than four thousands years: only for
$t\lsim 10^4$~yr the shock produces a CR spectrum that is relatively hard
in the relativistic energy range (see Fig. 4a). For $t\gsim 10^4$~yr it
becomes progressively steeper so that the CR number density at energy
$\epsilon=10$~TeV, which generates TeV-energy $\gamma$-rays, has already
at $t=3\times 10^4$~yr decreased by more than two orders of magnitude.
Due to this reason TeV-energy $\gamma$-rays at $t\gsim 10^4$~yr are
produced by CRs accelerated at the earlier stage $t\lsim 10^4$~yr, and
the contribution of freshly accelerated CRs to the $\gamma$-ray
production is negligible.  The $\gamma$-ray flux decreases with time for
$t\gsim 10^4$~yr because of adiabatic cooling of the CRs accelerated
during the previous stage.

Due to its high temperature, region (b) is characterized by a relatively
large value of the sound speed $c_s=188$~km/s. When it breaks into the
region (b) the shock has a velocity of about $V_s=4500$~km/s. Therefore
even the initial Mach number $M =24$ is not very high. Only during an
initial period $t\lsim 10^4$~yr the shock compression ratio $\sigma$ is
substantially larger than 4 (see Fig. 4b), and high energy CRs are
produced with relatively high efficiency. At later stages $t\gsim 10^4$~yr
the shock decelerates rapidly (Fig. 4d) and becomes too weak to accelerate
high energy CRs efficiently. Due to the small Mach number the total shock
compression ratio $\sigma$ at $t\gsim 10^4$~yr is less then 4 and even the
lower limit $q_{\rm t}$ for the power law index at $t\gsim 10^4$~yr
becomes greater than 5 (Fig. 4b). The CR spectrum $f_s\propto p^{-q}$
produced by the SN shock at this stage is very steep ($q\geq q_{\rm t}>5$)
which is tantamount to a low acceleration efficiency.

One can see from Fig. 4c that the number of high energy CRs with $p/{\rm
mc}\gsim 10^3$, produced in the SNR, increases only during the initial
period $t\lsim 10^4$~yr and then remains nearly constant during shock
propagation in the bubble. The lower energy particles, on the other hand,
continue to be accelerated beyond this time, steepening the overall
spectrum at lower energies, whereas $E_c$ remains nearly constant. At the
end of this stage about 22 percent of the explosion energy is transformed
into CRs (see Fig. 4f). Since the chemical composition of the bubble gas
is more and more modified towards a pure interstellar composition the
further out in radius the shock extends, we conclude from Fig. 4c that the
chemical composition of the CR spectrum becomes more wind material-like
(heavier) towards higher particle energies. The same tendency is seen in
the later SN II case (see Fig. 5c). This effect is peculiar to wind SNe
but it is only one of several factors which lead to an increasingly
heavier chemical composition of the Galactic CRs with increasing particle
energy.

Approximately, neglecting the mass in the wind, the behavior of CR and
$\gamma$-ray production in the bubble is determined by the dynamical scale
length $l_0$ and the corresponding time scale
\[
t_0=l_0/V_0,\mbox{  where  }V_0=\sqrt{2E_{sn}/M_{ej}}
\]
is the mean ejecta velocity.  As in the case of a uniform medium $l_0$ is
the length over which the amount of swept-up material equals the ejected
mass $M_{ej}$. The principal difference to the case of a uniform medium is
that in the case of the bubble the shock expansion starts from the initial
size $R_s\approx R_1$ which is much larger than $l_0$.

 Therefore the appropriate definition for the dynamic scale in the case
under consideration is
\begin{equation} 
l_0=\frac{M_{ej}}{4\pi R_1^2 \rho_b},  
\end{equation}
where $\rho_b$ is the bubble density.
It gives the value $l_0=4.92$~pc which is much smaller than
$R_1$.  The corresponding time scale is
$t_0=1522$~yr.  One can see from Fig. 3 that in agreement with
this time scale the TeV-energy $\gamma$-ray emission reaches its peak
value
$F_{\gamma}\approx 10^{-13}$~cm$^{-2}$s$^{-1}$ at $t\approx
t_1+2t_0\approx 10^4$~yr, that is  at the beginning of the Sedov phase
as in the case of a uniform ISM.  Later on the $\gamma$-ray flux
gradually
decreases (see Fig. 3) due to the decrease of the CRs production  because
of the
decrease of the shock strength (see Fig. 4b).

For a long time (at least up to $t=8\times 10^4$~yr) the CR energy content
$E_c$ remains nearly constant. In the case under consideration the
adiabatic cooling of CRs is less important compared with the uniform ISM
case because the shock size at the Sedov phase varies over a very small
range: during the period from $t\approx 7\times 10^3$~yr to $t\approx
7\times 10^4$~yr the shock size increases only by a factor of two (see
Fig. 4d).

At $t>10^4$~yr the CR energy content at the shock front is mainly in the
form of low energy particles because freshly accelerated CRs are
characterized by a progressively steeper spectrum (see Fig. 4a). Note that
the local peak (bump) in $p^4f_s(p)$ at $p\sim 10^5 mc$ represents the
so-called escaping particles (Berezhko 1986a; Berezhko et al. 1996). Fig.
4c shows that the overall CR spectrum $N(p,t)$ in the relativistic energy
range remains nearly constant for $t>10^4$~yr. It has an almost pure power
law form $N\propto p^{-2.1}$ in the momentum interval $10\lsim p/mc\lsim
10^5$ (Fig. 4c).

The efficiency of high energy $\gamma$-ray production increases again for
$t>2\times 10^4$~yr due to the increasing gas density (see Fig. 4e). But
even at $t\sim 10^5$~yr, when $N_g(R_s)\sim 1$~cm$^{-3}$ the TeV-energy
$\gamma$-ray flux remains significantly lower than
$10^{-12}$~cm$^{-2}$s$^{-1}$.

Note that during this stage the total number of high energy CRs remains almost
constant since the shock becomes too slow and weak, and does not accelerate
energetic particles any more. Nevertheless one can expect a significant
increase of high energy $\gamma$-ray production later on when the strong
population of previously produced CRs will expand outward and finally reach the
shell (c) and its boundary. During this stage an almost constant number of CRs
progressively interacts with an increasing amount of gas and this will lead to
an increasing $\gamma$-ray production. Unfortunately, it is not simple to
describe the dynamics of the system during this late stage. One has for example
to expect a strong increase of the CR diffusion coefficient due to the decrease
of the CR energy density as well as magnetohydrodynamic wave damping by ion
neutral friction in the partially ionized shell gas. This suggests a much
faster outward loss of CRs compared to the Bohm diffusion case, unless
the SNR is surrounded by a hot, fully ionized gas which reflects the
particle back across the shell.

In Fig. 3 we also present the time dependence of the TeV-energy
$\gamma$-ray flux for the same case of a SNR Ib but for a higher
(asymptotic) ISM number density $N_H=30$~cm$^{-3}$. In this case we have
$R_1=14.6$~pc, $R_2=34.7$~pc, $N_b=2.82\times10^{-2}$~cm$^{-3}$,
$P_b=1.1\times 10^{-11}$~dyne/cm$^2$, $B_b=16.7$~$\mu$G.  One can see that
the character of the SNR evolution is very similar to the previous case.
For $t<2544$~yr SN the shock propagates through the region (a) of the
supersonic WR-wind and its evolution is initially identical to the
previous case.

The bubble region, reached by the SNR shock at $t_1=2544$~yr, is characterized
by the length scale $l_0=1.8$~pc and the time scale $t_0=555$~yr. The value
$F_{\gamma}=3\times 10^{-13}$~cm$^{-2}$s$^{-1}$, reached at $t\approx
t_1+2t_0=3654$~yr, would be the peak flux for a uniform medium with number
density $N_0=N_b$. In contrast to the previous case, the $\gamma$-ray flux
continues to increase for $t>3654$~yr because at this period the SN shock
enters the region (c), where the gas density starts to increase rapidly (see
Fig. 4d).

One can see from Fig. 3 that the $\gamma$-ray flux reaches the value
$F_{\gamma}\approx 10^{-10}$~cm$^{-2}$s$^{-1}$ at $t\approx 3\times
10^4$~yr. A comparison with the previous case shows that this peak flux
scales as $N_H$ as in the case of a uniform ISM (e.g. Berezhko \& V\"olk
1997).  However, it is essentially lower (at least by a factor of 100)
than in the case where the ISM is uniform with the same density $N_H$.  
The main reason of this low efficiency of $\gamma$-ray production is that
in the case under discussion the majority of CRs is produced in the low
density bubble. When the SN shock enters the high density shell region (c)
it becomes weak and therefore an inefficient CR accelerator. In this
phase, in comparison with the case of a uniform ISM, the same amount of
CRs mainly produced at previous evolutionary phases generates a much lower
$\gamma$-ray flux because of the poor spatial overlap between the CR and
gas distributions (see Berezhko \& V\"olk 1997). The peak of the gas
density distribution lies just behind the shock front, whereas the CRs are
sitting deeper inside where the density of the gas is much lower.

\subsection{Type II supernova} 
We model the type II SN case as a progenitor
star with initial mass $15M_{\odot}$ 
(e.g. Leitherer et al. 1992)
that spends a time period $\Delta
t_w=4\times 10^6$~yr on the main-sequence with a mass-loss rate
$\dot{M}=2.5\times 10^{-7}M_{\odot}$~yr$^{-1}$ and wind velocity
$V_w=2000$~km/s, and then the time $\Delta t_w=10^5$~yr in the RSG phase with
mass-loss rate $\dot{M}=2\times 10^{-5}M_{\odot}$~yr$^{-1}$ and wind velocity
$V_w=15$~km/s.  The MS wind creates a bubble of size $R_2=61$~pc in the ISM
with $N_H=0.3$~cm$^{-3}$. According to Eqs. (14) and (15) the bubble is
characterized by $N_b=6.6\times10^{-3}$~cm$^{-3}$ and $P_b=4.4\times
10^{-13}$~dyne/cm$^2$.

The RSG wind occupies the region of size $R_f=V_w \Delta t_w=1.54$~pc. At
this point the ram pressure $\rho V_w^2$ of RSG wind exceeds the thermal
pressure $P_b$ in the bubble. Therefore we neglect the shell which can be
formed in the RSG wind due to its interaction with the ambient bubble
material.  We model the transition zone between the RSG wind and the
bubble by the set of parameters $\rho=\rho(R_f)(R_f/r)^{3.5}$,
$V_w=V_w(R_f)(R_f/r)^2$. We introduce this zone to match smoothly the gas
densities $N_g$ between regions (a) and (b), see Fig. 1. It contains a
small amount of gas and plays no role in the overall SNR evolution.

We use a magnetic field strength $B=2\times10^{-4}$~G in the RSG wind at the
distance $r=10^{17}$~cm. It formally correspond to $B_*=1$~G,
$R_*=3\times10^{13}$~cm and $\Omega=3\times 10^{-8}$. Our B-field is about
60 times smaller than that assumed by V\"olk \& Biermann (1988). Their
magnetic field implies an Alfv\'en velocity $c_{\rm a}$ which is
considerably larger than the wind speed $V_w$. This renders the existence
of the RSG mass outflow as a supersonic wind problematic (Axford 1994). In
our case the RSG wind is superalfv\'enic with $c_{\rm a}\approx 0.5V_w$.
It might be possible to model a slow outflow starting with a much higher
stellar magnetic field which is then subalfv\'enic. However, we shall not
attempt such a dynamical construction in this paper.

We start the computation of the SNR evolution with the initial conditions
$t_i=3.17$~yr, $R_{pi}=10^{17}$~cm, $V_{pi}=10^4$~km/s.

Fig. 5 illustrates SN shock evolution, CR and $\gamma$-ray production for
an ISM with $N_H=30$~cm$^{-3}$. The character of the CR spectrum
$f_s(p,t)$, shown in Fig. 5a, and of the overall CR spectrum $N(p,t)$,
plotted in Fig. 5c, are similar to the previous case of the SN Ib for
$t<500$~yr when the SN shock propagates through the region (a) of the
supersonic wind. The value of the cutoff momentum $p_m\approx
2\times10^5$~mc is consistent with formula (28). The acceleration
efficiency and the corresponding shock modification are very high. During
the initial stage $t\approx 10$~yr the shock compression ratio reaches the
value $\sigma=15$ and then slowly decreases due to the shock deceleration
(see Fig. 5b).  As a consequence the high energy part of the CR spectrum
is
extremely hard $N\propto p^{-1.2}$ at $10^3\lsim p/mc\lsim 10^5$.

\begin{figure*}
\resizebox{\hsize}{!}{\includegraphics{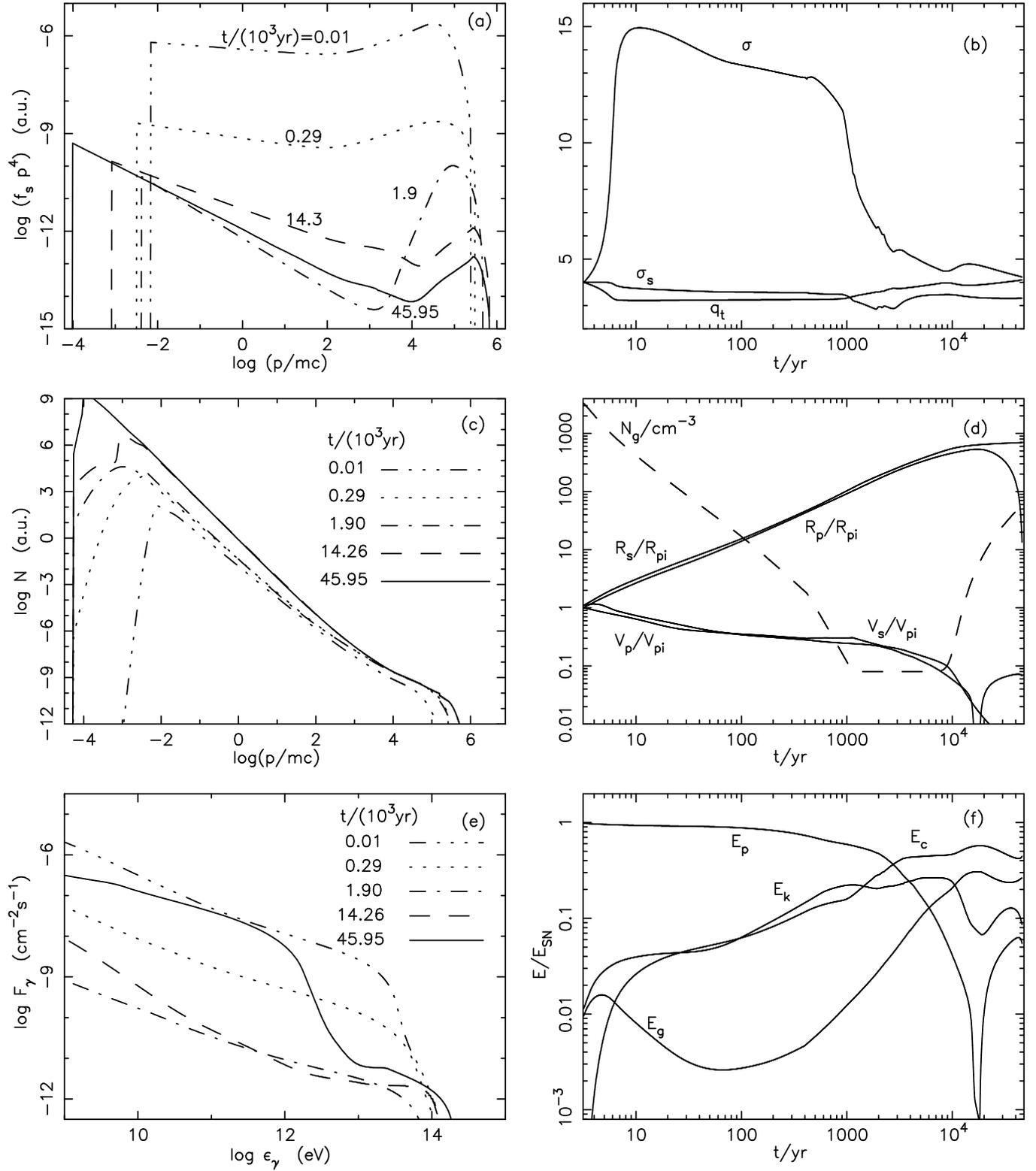}}
\caption{The same as Fig. 2 but for the case of the remnant of a type II
SN in an ISM with the hydrogen number density $N_H=30$~cm$^{-3}$, and
encompassing the entire SNR evolution. The normalization parameters are,
respectively, given by $R_{pi}=10^{17}$~cm, $V_{pi}=10^4$~km/s}
\end{figure*}

It is interesting to note that the $\gamma$-ray spectrum
$F_{\gamma}\propto \epsilon_{\gamma}^{-0.5}$ (Fig. 5e) is much steeper at
$10^{11}\lsim \epsilon_{\gamma}\lsim 10^{13}$~eV than the integral CR
spectrum:  the CRs in the upstream region $r>R_s$, which make the integral
CR spectrum extremely hard, play a much less important role in the
formation of the $\gamma$-ray spectrum, because they occupy a relatively
lower density region.

In the present case the $\gamma$-ray generation within the ejecta is
negligible. Therefore, according to relation (30), the expected
$\gamma$-ray flux $F_{\gamma}$(1 TeV) should be about $5\times 10^3$ times
larger than the flux $F_{\gamma}'$(1 TeV) in the case of the WR wind. One
can see from Fig.3 and Fig.5e, that the calculated $\gamma$-ray flux can
be represented in the form (31) with $F_{10}=1.2\times
10^{-8}$cm$^{-2}$s$^{-1}$ in agreement with relation (30).

During these initial 500 years the SN shock sweeps up $M_{sw}=2M_{\odot}$,
CRs absorb about 10\% of the explosion energy (see Fig. 5f) and
the expected $\gamma$-ray flux $F_{\gamma}$(1 TeV) exceeds the value
$10^{-10}$~cm$^{-2}$s$^{-1}$ (see Fig. 3).

During the period $500<t<1000$~yr the SN shock propagates through the
intermediate zone (see Fig. 5d) which contains only a small amount of
matter. The CR energy content slowly increases, but the CR spectrum
becomes progressively steeper. Together with the decrease of the gas
density this leads to a decrease of the $\gamma$-ray flux
$F_{\gamma}\propto t^{-2}$ (see Fig.3).

For $t>10^3$~yr the SN shock propagates through the bubble whose
characteristics are similar to the bubble around the progenitor of a type
Ib SN, in an ISM with the same density $N_H$ (see Fig. 1).  Therefore at
this stage the expected $\gamma$-ray production continues to decrease up
to the time $t=4\times 10^3$~yr, when the $\gamma$-ray production by CRs
accelerated during previous stages drops to a level that corresponds to a
flux $F_{\gamma}\sim 10^{-12}$~cm$^{-2}$s$^{-1}$ typical for a uniform
medium with $N_0=N_b$. For $t>4\times 10^3$~yr $\gamma$-rays produced in
the bubble material start to dominate and the $\gamma$-ray flux increases
with time. At $t=10^4$~yr the SN shock reaches the shell region (c) where
the density increases. Therefore the expected $\gamma$-ray flux increases
more rapidly for $t>10^4$~yr (see Fig. 3), even though the CR production
becomes quite low in this phase (see Fig. 5f). More than 40\% of the
explosion energy goes into CRs during the full SNR evolution.

In Fig. 3 we present also the calculated $\gamma$-ray flux $F_{\gamma}$(1
TeV) for the case $N_H=0.3$~cm$^{-3}$. The time-dependence of
$F_{\gamma}$(1 TeV) for $t<2\times 10^3$~yr is identical to the previous
case. For $t>2\times 10^3$~yr the SN shock propagates through the bubble
whose characteristics are now similar to the case of a type Ib SNR with
$N_H=0.3$~cm$^{-3}$. Therefore, the expected $\gamma$-ray production
during late phases $t>10^4$~yr is close to that case (see Fig. 3).

\section{Summary}

Our numerical results show that when a SN explodes into a circumstellar
medium strongly modified by a wind from a massive progenitor star, then
CRs are accelerated in the SNR almost as effectively as in the case of a
uniform ISM (Berezhko et al. 1994, 1995, 1996; Berezhko \& V\"olk 1997):
about $20\div 40$\% of the SN explosion energy is transformed into CRs
during the active SNR evolution.

During SN shock propagation in the supersonic wind region very soon the
acceleration process reaches a quasi-stationary level which is
characterized by a high efficiency and a correspondingly large
shock modification. Despite the fact that the shock modification is much
stronger than predicted by a two-fluid hydrodynamical model (Jones \& Kang
1992), the shock never becomes completely smoothed by CR backreaction: a
relatively strong subshock always exists and plays an important dynamical
role. As in the case of a uniform ISM, the spurious complete shock
smoothing in hydrodynamic models is the result of an underestimate of the
role of geometric factors.

Due to the relatively small mass contained in the supersonic wind region
CRs absorb there only a small fraction of the explosion energy (about 1\%
in the case of a SN type Ib, and 10\% in the case of a SN type II) and the
SNR is still very far from the Sedov phase after having swept up this
region. Therefore we conclude, that the CRs produced in this region
should not play a very significant role for the formation of the observed
Galactic CR energy spectrum.

The peak value of the CR energy content in the SNR is reached when the SN
shock sweeps up an amount of mass roughly equal to several times the
ejected mass. This takes place during the SN shock propagation in the
modified bubble. In a purely adiabatic bubble the sweep-up would occur at
the beginning of the shell. Compared with the uniform ISM case the
subsequent adiabatic CR deceleration is less important in the case of a
modified circumstellar medium. The main amount of CRs in this case is
produced when the SN shock propagates through the bubble. In this stage
the dynamical scale length is much smaller than the shock size. Therefore
the relative increase of the shock radius during the late evolution stage
and the corresponding adiabatic effects are small. These configurational
properties lead to potentially interesting changes of the CR chemical
composition with particle energy (see Sect. 3.1)

In the case of the modified circumstellar medium the CR and $\gamma$-ray
spectra are more variable during the SN shock evolution than in the case
of a uniform ISM (Berezhko \& V\"olk 1997). At the same time the form of
the resulting overall CR spectrum is rather insensitive to the parameters
of the ISM as in the case of uniform ISM.  The reason is that the main
amount of CRs are produced in the latest phase which has still a strong
enough shock. Roughly speaking, the overall CR spectrum (except the most
energetic CRs) is mainly formed at the stage when the shock compression
ratio lies between 4 and 5.

The maximum energy of the accelerated CRs reached during the SNR evolution is
about $10^{14}$~eV for protons in all the cases considered.

Our results confirm the important conclusion, reached for the case of a uniform
ISM before, that the diffusive acceleration of CRs in SNRs is able to generate
the observed CR spectrum up to an energy $\sim 10^{14}$~eV, if the CR diffusion
coefficient is as small as Bohm limit. This disregards the possibility of
turbulent field amplification, as discussed in Sect. 2.

In the case of a SN Ib the expected TeV-energy $\gamma$-ray flux,
normalized to a distance of 1 kpc, remains lower than
$10^{-12}$~cm$^{-2}$s$^{-1}$ during the entire SNR evolution if the ISM
number density is less than 1~cm$^{-3}$ except for an initial short period
$t<100$~yr when it is about $10^{-11}$~cm$^{-2}$s$^{-1}$.  Only for a
relatively dense ISM with $N_H=30$~cm$^{-3}$ the expected $\gamma$-ray
flux is about $10^{-10}$~cm$^{-2}$s$^{-1}$ at late phases $t>10^4$~yr.
A similar situation exists at late phases of SNR evolution in
the case of SN II. It is interesting to note that the expected
$\gamma$-ray flux is considerably lower, at least by a factor of
hundred, compared with the case of uniform ISM of the same density $N_H$.
This confirms the preliminary result reported earlier (Berezhko \& V\"olk
1995)

The type II SN explodes into the dense wind of the red supergiant progenitor
star. During the first several hundred years $t_m$ after the explosion, the
expected TeV-energy $\gamma$-ray flux at a distance $d=1$~kpc exceeds the value
$10^{-9}$~cm$^{-2}$s$^{-1}$ and can be detected up to the distance $d_m=30$~kpc
with present instruments like HEGRA, Whipple or CAT. This distance is of the
order of the diameter of the Galactic disk (see also Kirk et al. 1995).  
Therefore all Galactic SNRs of this type whose number is $N_{sn}=\nu_{sn}t_m$
should be visible. But in this case we can expect at best $N_{sn}\sim 10$ such
$\gamma$-ray sources at any given time.

The typical value of the cutoff energy of the expected $\gamma$-ray flux
is about $10^{13}$~eV, if the CR diffusion coefficient is as small as the
Bohm limit. In this respect the negative result of high-threshold arrays
(Borione et al. 1995; Allen, G.E. et al. 1995; Allen, N.H. et al. 1995) in
searching of $\gamma$-ray emission from Galactic SNRs is not surprising
because their threshold $E_{th}\sim 50$~TeV exceeds the cutoff energy of
the expected $\gamma$-ray flux; marginally this also holds for the
negative results of the lower threshold $E_{th}>$~20 TeV AIROBICC array
(Prahl \& Prosch 1997; Prosch et al. 1996). It is less obvious how to
interpret the negative results of imaging atmospheric Cherenkov telescopes
with thresholds less than about 1 TeV (Mori et al. 1995; Lessard et al.
1997; Hess et al. 1997; Hess 1998; Buckley et al. 1998). For core collapse
SN of types II or Ib with quite massive progenitors one can in part
explain this fact by the extremely low $\pi^0$-decay $\gamma$-ray
intensity expected from such SNRs during the period of SN shock
propagation through the low-density hot bubble. An alternative possibility
relates to the assumption of the Bohm limit for the CR diffusion
coefficient which can be too optimistic, in particular for the
quai-perpendicular geometry in wind-blown regions from rotating stars. For
a slightly more general discussion of SNR $\gamma$-rays in stellar wind
cavities, see V\"olk (1997).

%
%

\begin{acknowledgements}
This work has been supported in part by the Russian Foundation of Basic
Research grant 97-02-16132. One of the authors (EGB) gratefully
acknowledges the hospitality of the Max-Planck-Institut f\"ur Kernphysik
where part of this work was carried out under grant 05 3HD76A 0 of the
Verbundforschung A\&A of the German BMBF.
\end{acknowledgements}

\appendix
\section{Similarity solution}
Consider a shock of radius $R_s=V_st$  that expands with constant speed $V_s$
into the wind region, whose parameters are described by expressions (12), (13).
We introduce the similarity variables
\[x=r/R_s ,\]
\[f(r,p,t)=\Phi (x,p)/t^2,\]
\[P_c(r,t)=\Pi _c(x)/t^2,\]
\[w(r,t)=W(x),\]
\[ \rho(r,t)=\Omega (x)/t^2,\]
\[P_g(r,t)=\Pi _g(x)/t^2.\]
Then the diffusive transport equation for the CR distribution function, eq.(1),
 and the gas dynamic equations (2)-(4) can be written in the form
\[
2\Phi +x {\partial \Phi \over \partial x}=
{1\over x^2V_s}{\partial\over\partial x}
x^2 K {\partial \Phi \over\partial x}-\frac{W}{V_s}{\partial \Phi \over\partial x}+
\]
\[
{1\over x^2V_s}{\partial\over\partial x}(x^2W){p\over 3}
{\partial \Phi \over\partial p}+\frac{\eta (V_s-W_1)\Omega_1}{4\pi m p_{inj}^3 V_s}
\delta (x-1),
\]
\[
{\partial\over\partial x}[(xV_s-W)\Omega]-\frac{2W\Omega}{x}-3V_s\Omega
=0,
\]
\[
(xV_s-W)\Omega {\partial W \over \partial x} =
{\partial\over\partial x}(\Pi_{\rm g}+\Pi_{\rm c}),
\]
\[
(W-xV_s){\partial \Pi_{\rm g} \over \partial x}+{\gamma_{\rm g}\over x^2}{\partial\over\partial x}
(x^2W)\Pi_{\rm g}= \alpha_a (\gamma_{\rm g}-1)
c_{\rm a}{\partial \Pi_{\rm c}\over \partial x} ,
\]

taking into account that in the wind region the Alfv\'en speed $c_{\rm a}$ is
constant and small compared with the shock speed $V_s$, and plausibly assume
that the CR diffusion coefficient has the form $\kappa (r,p,t)=K(x,p)R_s$. The
boundary conditions do not contain the time explicitly. Therefore the above
similarity solution is appropriate. The only factor which violates these
assumptions it is the initial condition for CRs, which contains the time $t=t_i$
in explicit form. Therefore the exact solution will deviate from the similarity
solution only during some short initial period of several $t_i$, as long as the
shock speed is constant.


\begin{thebibliography}{}

\bibitem[1995]{bbb} 
Allen G.E., Berley D., Biller S. et al., 1995 Proc.
24th ICRC (Roma) 2, 443 
\bibitem[1995]{bu} 
Allen N.H., Bond L.A., Budding E., et al., 1995 Proc. 24th ICRC (Roma)
2, 447
\bibitem[1996]{atah} 
Atoyan A.M., Aharonian F.A. 1996, MNRAS 278, 525
\bibitem[1994]{axf} 
Axford W.I. 1994, ApJS 90, 937 
\bibitem[1977]{axf}
Axford W.I., Leer E., Skadron G. 1977, Proc. 15th ICRC (Plovdiv) 11, 132
\bibitem[1997]{bar} 
Baring M.G., Ellison D.C., Reynolds S.P., Grenier I.A., Goret P.
1999, ApJ 513, 311 
\bibitem[1978]{bel} 
Bell A.R. 1978, MNRAS 182, 147; MNRAS 182, 443 
\bibitem[1995]{bee}
Bennett L., Ellison D.C. 1995, J. Geophys. Res. 100, A3, 34439
\bibitem[1986]{bere86a} 
Berezhko E.G. 1986a, Proc. Int. School \& Workshop on Plasma Astrophysics
(Sukhumi) (ESA SP-251-ISSN 0379-6566) 271
\bibitem[1986]{bere86b} 
Berezhko E.G. 1986b, Pis'ma Astron. Zh. 12, 842 
\bibitem[1996]{egb} 
Berezhko E.G. 1996, Astropart. Phys. 5, 367 
\bibitem[1999]{berela} 
Berezhko E.G., Ellison D.C. 1999, ApJ, 526, 385
\bibitem[1999]{berelb} 
Berezhko E.G., Ellison D.C. 2000, ApJ, to be published 
\bibitem[1988]{ber}
Berezhko E.G., Krymsky G.F. 1988, Soviet Phys. Uspekhi. 12, 155
\bibitem[1995]{bv}
Berezhko E.G., V\"olk H.J. 1995, Proc. 24th ICRC (Roma) 3, 380
\bibitem[1997]{ber} 
Berezhko E.G., V\"olk H.J. 1997, Astropart. Phys. 7, 183 
\bibitem[1990]{bkt} 
Berezhko E.G., Krymsky G.F., Turpanov A.A. 1990, Proc. 21st ICRC
(Adelaide) 4, 101 
\bibitem[1994]{ber2} 
Berezhko E.G., Yelshin V.K., Ksenofontov L.T. 1994, Astropart. Phys. 2,
215
\bibitem[1995]{ber3} 
Berezhko E.G., Ksenofontov L.T., Yelshin V.K. 1995, Nuclear Phys. B
(Proc.Suppl.) 39A, 171 
\bibitem[1996]{ber4}
Berezhko E.G., Yelshin V.K., Ksenofontov L.T. 1996, JETP 82, 1
\bibitem[1993]{bier1} 
\bibitem[1989]{vsb}
Berezinsky V.S., Ptuskin V.S., 1989, Sov. Astron. Lett. 14, 304
\bibitem[1990]{bbg}
Berezinsky V.S., Bulanov S.V., Ginzburg V.L., Dogiel V.A., Ptuskin
V.S. 1990, in {\it Astrophysics of Cosmic Rays}, Horth-Holland Elsevier
Science Publ. B.V., Amsterdam
Biermann P.L. 1993, A\&A 271, 649; 1993, A\&A 277, 691 
\bibitem[1987]{blan} 
Blandford R.D., Eichler D. 1987, Phys. Rept. 154, 1 
\bibitem[1978]{bo} 
Blandford R.D., Ostriker J.P. 1978, ApJ 221, L29 
\bibitem[1995]{bri} 
Borione A., Catanese M., Covault C.E., et al., 1995 Proc. 24th ICRC
(Roma) 2, 439 
\bibitem[1998]{buck} 
Buckley J.H., Akerlof C.W., Carter-Lewis D.A., et al. 1998, A\&A 329,
639
\bibitem[1982]{che} 
Chevalier R.A. 1982, ApJ 259, 302
\bibitem[1989]{chel} 
Chevalier R.A., Liang E.P. 1989, ApJ 344, 332
\bibitem[1993]{chuv} 
Chuvilgin L.G., Ptuskin V.S. 1993, A\&A 279, 278
\bibitem[1992]{dejh} 
de Jager O.C., Harding A.K. 1992, ApJ 396, 161
\bibitem[1990]{dorf90} 
Dorfi E.A. 1990, A\&A 234, 419
\bibitem[1991]{dorf91} 
Dorfi E.A. 1991, A\&A 251, 597 
\bibitem[1983]{dru}
Drury L.O'C. 1983, Rep. Prog. Phys. 46, 973 
\bibitem[1984]{drur84}
Drury L.O'C. 1984, Adv. Space Res. 4, 185 
\bibitem[1989]{dmv} 
Drury L.O'C, Markiewicz W.J., V\"olk H.J. 1989, A\&A 225, 179 
\bibitem[1994]{dru}
Drury L.O'C., Aharonian F.A., V\"olk H.J. 1994, A\&A 287, 959
\bibitem[1995]{druvb} 
Drury L.O'C., V\"olk H.J., Berezhko E.G. 1995, A\&A 299, 222 
\bibitem[1994]{ellet} 
Ellison D.C., Reynolds S.P., Borkowsky K., et al., 1994, PASP 106, 780
\bibitem[1995]{ell}
Ellison D.C., Baring M.G., Jones F.C. 1995, ApJ 453, 873
\bibitem[1996]{g-s} 
Garcia-Segura G., Langer N., Mac Low M.-M. 1996, A\&A 316, 133 
\bibitem[1993]{giac} 
Giacalone J., Burgess D., Schwartz S.J., Ellison D.C. 1993, ApJ 402,
550 
\bibitem[1997]{hes}
Hess M., for the HEGRA coll'n 1997, Proc. 25th ICRC (Durban) 3, 229
\bibitem[1998]{hess}
Hess M. 1998, PhD Thesis Univ. Heidelberg
\bibitem[1987]{jok} 
Jokipii J.R. 1997, ApJ 313, 842 
\bibitem[1992]{jonk}
Jones T.W., Kang H. 1992, ApJ 396, 575
\bibitem[1981]{jon} 
Jones E.M., Smith B.W., Straka W.C. 1981, ApJ 249, 185 
\bibitem[1989]{kahn} 
Kahn F.D., Breitschwerdt D. 1989, MNRAS 242, 209 
\bibitem[1991]{kah} 
Kang H., Jones T.W. 1991, MNRAS 249, 439 
\bibitem[1995]{jgk} 
Kirk J.G., Duffy P., Ball L. 1995, A\&A 293, L37 
\bibitem[1996]{jgket} 
Kirk J.G., Duffy P., Gallant Y.A. 1996, A\&A 314, 1010 
\bibitem[1964]{kry} 
Krymsky G.F. 1964, Geomag. Aeron. 4, 977
\bibitem[1977]{krym} 
Krymsky G.F. 1977, Soviet Phys. Dokl. 23, 327
\bibitem[1983]{lagc} 
Lagage P.O., Cesarsky C.J. 1983, A\&A 125, 249
\bibitem[1982]{leem82} 
Le, M.A. 1982, J.Geophys. Res. 87, 5063
\bibitem[1992]{leith} 
Leitherer C., Robert C., Durissen L. 1992, ApJ 401, 596 
\bibitem[1997]{less} 
Lessard R.W., Boyle P.J., Bradbury S.M., et al., 1997 Proc 25th ICRC
(Durban) 3, 233 
\bibitem[1991]{los}
Losinskaya T.A. 1991, Proc. 22th ICRC (Dublin) 5, 123 
\bibitem[2000]{lub}
Lucek S.G. \& Bell A.R., 2000, Mon. Not. R. Astron. Soc., in press
\bibitem[2000]{blu}
Bell A.R., Lucek S.G., 2000, Mon. Not. R. Astron. Soc., submitted
\bibitem[1997]{malk}
Malkov M.A. 1997, ApJ 485, 638 
\bibitem[1995]{malkv95} 
Malkov M.A., V\"olk H.J. 1995, A\&A 300, 605 
\bibitem[1995]{malkv96} 
Malkov M.A., V\"olk H.J. 1996, Adv. Space Res. 21, No. 4, 551 
\bibitem[1990]{mdv}
Markiewicz W.J., Drury L.O'C., V\"olk H.J. 1990, A\&A 236, 487
\bibitem[1996]{mast} 
Mastichiadis A. 1996, A\&A 305, L53
\bibitem[1996]{mastdj} 
Mastichiadis A., de Jager O.C. 1996, ApJ, 311, L5
\bibitem[1982]{mckv} 
McKenzie J.F., V\"olk H.J. 1982, A\&A 116, 191
\bibitem[1995]{mh} 
Mori M., Hara T., Hayashida N., et al., 1995 Proc 24th ICRC (Roma)
2,487 
\bibitem[1988]{ostr} 
Ostrowski M. 1988, MNRAS 233, 257 
\bibitem[1965]{park} 
Parker E.N. 1965, Planet. Space Sci. 13, 9
\bibitem[1997]{pra} 
Prahl J., Prosch C. 1997 Proc 25th ICRC (Durban) 3, 217 
\bibitem[1995]{pro} 
Prosch C., Feigl E., Karle A., et al., 1996, A\&A 314, 275 
\bibitem[1988]{ques88} 
Quest K.B. 1988, JGR 93, 9649
\bibitem[1998]{rey} 
Reynolds S.P. 1998, ApJ 493, 375
\bibitem[1990]{shi} 
Shigeyama T., Nomoto K. 1990, ApJ 360, 242
\bibitem[1991]{trat}
Trattner K.J., Scholer M., 1991, Geophys. Res. Lett. 18, 1817
\bibitem[1994]{kjt} 
Trattner K.J., M\"obius E., Scholer M. et al., 1994, JGR 99, 389 
\bibitem[1984]{vo}
V\"olk H.J., 1984, in Proc. 19th Recontre de Moriond Astrophysics
Meeting: "High Energy Astrophysics" (ed. J. Tran Thanh Van),
editions Frontieres, Gif-sur-Yvette, France, p. 281 ff
\bibitem[1997]{vlk} 
V\"olk H.J. 1997, in: Towards a Major Atmospheric Cherenkov Detector - V,
O.C. de Jager (ed.), Kruger Park, p.87 ff
\bibitem[1988]{qvlk}  
V\"olk H.J., Biermann P.L. 1988, ApJ 333, L65  
\bibitem[1984]{vdm} 
V\"olk H.J., Drury L. O'C., McKenzie J.F., 1984, A\&A 130, 19 
\bibitem[1977]{waew} 
Weaver R., McCray R., Castor J. et al., 1977, ApJ 218, 377

\end{thebibliography}
\end{document}